\documentclass{aa}  

\usepackage{graphicx}
\usepackage{txfonts}
\usepackage[dvipsnames]{xcolor}
\usepackage[figuresright]{rotating}
\usepackage{hyperref}
\hypersetup{
    colorlinks=true,
    allcolors=blue
    }
\begin{document} 
\title{Toward fully compressible numerical simulations of stellar magneto-convection with the RAMSES code}
\author{
    J.~R.~Canivete Cuissa\inst{1,2}
    \and
    R.~Teyssier\inst{2,3}
}
\institute{
    IRSOL Istituto Ricerche Solari ``Aldo e Cele Daccò'' Locarno, Università della Svizzera Italiana (USI), \\
    Via Patocchi 57 -- Prato Pernice, 6605 Locarno-Monti, Switzerland\\
    \email{jose.canivete@irsol.usi.ch}
 \and 
    Center for Theoretical Astrophysics and Cosmology, Institute for Computational Science (ICS),\\University of Zurich, Winterthurerstrasse 190, 8057 Z{\"u}rich, Switzerland 
 \and
    Department of Astrophysical Sciences, Princeton University, 4 Ivy Lane, Princeton, New Jersey, 08544, United States
}
\date{Received 26 November 2021 / Accepted 8 June 2022}
\abstract
    {Numerical simulations of magneto-convection have greatly expanded our understanding of stellar interiors and stellar magnetism. Recently, fully compressible hydrodynamical simulations of full-star models have demonstrated the feasibility of studying the excitation and propagation of pressure and internal gravity waves in stellar interiors, which would allow for a direct comparison with asteroseismological measurements. However, the impact of magnetic fields on such waves has not been taken into account yet in three-dimensional simulations.}
    {We conduct a proof of concept for the realization of three-dimensional, fully compressible, magneto-hydrodynamical numerical simulations of stellar interiors with the RAMSES code. }
    {We adapted the RAMSES code to deal with highly subsonic turbulence, typical of stellar convection, by implementing a well-balanced scheme in the numerical solver. We then ran and analyzed three-dimensional hydrodynamical and magneto-hydrodynamical simulations with different resolutions of a plane-parallel convective envelope on a Cartesian grid. }
    {Both hydrodynamical and magneto-hydrodynamical simulations develop a quasi-steady, turbulent convection layer from random density perturbations introduced over the initial profiles. The convective flows are characterized by small-amplitude fluctuations around the hydrodynamical equilibrium of the stellar interior, which is preserved over the whole simulation time. 
    Using our compressible well-balanced scheme, we were able to model flows with Mach numbers as low as $\mathcal{M} \sim 10^{-3}$, but even lower Mach number flows are possible in principle.
    In the magneto-hydrodynamical runs, we observe an exponential growth of magnetic energy consistent with the action of a small-scale dynamo. 
    The weak seed magnetic fields are amplified to mean strengths of $37\%$ relative to the kinetic equipartition value in the highest resolution simulation. Since we chose a compressible approach, we see imprints of pressure and internal gravity waves propagating in the stable regions above and beneath the convection zone. In the magneto-hydrodynamical case, we measured a deficit in acoustic and internal gravity wave power with respect to the purely hydrodynamical counterpart of $16\%$ and $13\%$, respectively.
    }
    {
    The well-balanced scheme implemented in RAMSES allowed us to accurately simulate the small-amplitude, turbulent fluctuations of stellar (magneto-)convection.
    The qualitative properties of the convective flows, magnetic fields, and excited waves are in agreement with previous studies in the literature. The power spectra, profiles, and probability density functions of the main quantities converge with resolution. Therefore, we consider the proof of concept to be successful. 
    The deficit of acoustic power in the magneto-hydrodynamical simulation shows that magnetic fields must be included in the study of pressure waves in stellar interiors. 
    We conclude by discussing future developments.}
\titlerunning{Toward fully compressible MHD simulations of stellar convection with RAMSES}
\maketitle
%
%
\section{Introduction} 
\label{sec:introduction}
Thermal turbulent convection is one of the fundamental processes in stellar physics. It is responsible for the outward transport of the energy generated in the core, but it also affects the structure, the dynamics, and the evolution of the star. In our Sun, for example, the convective envelope shapes and influences the observable features and activity on the solar surface and in the overlying atmosphere \citep[][]{2009LRSP....6....2N, 2012LRSP....9....4S}. 
Core convection in massive stars might impact the star's lifetime by bringing fresh fuel into the core as the convective cells overshoot into the stable radiative zones above it \citep[][]{2017RSOS....470192S}, 
and convective mixing in asymptotic giant branch (AGB) stars provides a rich environment for nucleosynthesis \citep{2005ARA&A..43..435H}. 

Stellar magnetism is also linked to turbulent convective motions in stellar interiors. A small-scale dynamo operating in the near-surface turbulent convection zone of the Sun is possibly at the origin of the small-scale magnetic fields permeating the quiet photosphere \citep[][]{2014PASJ...66S...4L, 2018ApJ...859..161R}, 
while the cyclic regeneration of the solar large-scale magnetic field probably stems from the interplay between deep-convection zone turbulence, differential rotation, and magnetic flux transport \citep[][]{2013SAAS...39..187C, 2020LRSP...17....4C}. 
Moreover, convective cores in A- and B-type stars are likely able to develop a dynamo action \citep[][]{2005ApJ...629..461B, 2016ApJ...829...92A} 
and all low-mass stars appear to host dynamo-produced, small-scale, surface magnetic fields \citep[][]{2014IAUS..302....1L}. 

Finally, turbulent convective envelopes and cores also cause the excitation and propagation of a rich spectrum of oscillation modes in stellar interiors \citep[][]{2015LRSP...12....8H}. 
With the advent of asteroseismology, it is possible to determine properties of stellar internal structures from the observation of global oscillations across the Herzsprung-Russell diagram \citep[][]{2017A&ARv..25....1H, 2020FrASS...7...44B,2021RvMP...93a5001A}. 
It is therefore crucial to understand and characterize the formation and propagation of these modes, in particular in the presence of magnetic fields.

The classical modeling of convection is based on mixing length theory \citep[MLT,][]{1958ZA.....46..108B}. 
However, large-scale (magneto-)hydrodynamical  simulations of solar and stellar convection are nowadays employed to include multidimensional dynamical processes such as overshooting, oscillations, and dynamo action \citep[see, e.g.,][]{2002AN....323..213F, 2017A&A...600A.137F, 2004ApJ...601..512B, 2006ApJ...648L.157B, 2004ApJ...614.1073B, 2006ApJ...642.1057H, 2007ApJ...667..448M, 2008ApJ...673..557M, 2009ApJ...702.1078B, 2010ApJ...715L.133G, 2014ApJ...789..132R, 2014ApJ...786...24H, 2015ApJ...803...42H, 2015A&A...581A..42B, 2015ApJ...812...19T, 2016ApJ...829...92A, 2017A&A...599A...4K, 2018A&A...614A..78S, 2019ApJ...876....4E, 2020MNRAS.491..972A, 2020A&A...641A..18H, 2021A&A...653A..55H, 2021A&A...651A..66K}. 

Such simulations can be challenging for common numerical techniques as stellar convective flows are often highly subsonic and characterized by small perturbations around a hydrostatic equilibrium. 
In order to alleviate the time-stepping constraints imposed by the low-Mach number regime, numerical schemes solving the Navier-Stokes (or MHD) equations in the anelastic approximation are typically favored \citep[see][for a review]{2017LRCA....3....1K}. 

Although internal gravity waves are preserved in this approach, the physics of pressure waves are precluded and an artificial viscosity is required to achieve numerical stability.
Fully compressible simulations are therefore necessary to examine the properties and the dynamics of the complete spectrum of excited waves in stellar interiors, as well as the coupling between these modes \citep[][]{2011Sci...332..205B}. 
The computational cost of these simulations is higher than the anelastic ones, but \citet{2020A&A...641A..18H, 2021A&A...653A..55H} 
recently demonstrated the feasibility of this approach in two and three dimensions.

The next logical step is to extend the fully compressible framework to magnetized stars. Such simulations would allow to study the interaction between convective turbulent motions, magnetic fields, and global oscillations. In particular, it would enable a numerical analysis of the imprints of magnetic fields in global oscillations, which could be used to assess the presence of strong magnetic fields hidden in stellar interiors from asteroseismologic measurements \citep[][]{2015Sci...350..423F, 2020MNRAS.496..620G}. 

Therefore, we test the feasibility of three-dimensional, fully compressible, magneto-hydrodynamical numerical simulations of stellar convection. This work is hence a proof of concept: we intend to present and validate our approach by performing a convergence study and comparing qualitatively the results to previous works on stellar convection, dynamo action, and wave propagation. The physics and the numerical setup are therefore intentionally simplistic, and they will be enhanced and addressed in future works.

This paper is organized as follows: in Sect.\,\ref{sec:numerics} we briefly describe the code, the numerical techniques, and define the initial conditions of our simulations. The results are presented and discussed in Sect.\,\ref{sec:results}, while in Sect.\,\ref{sec:conclusion} we summarize the main aspects of our work and we give an outlook for future developments. 
%
%
\section{Numerics} 
\label{sec:numerics}
We ran our numerical simulations with the Adaptive Mesh Refinement (AMR) code RAMSES \citep{2002A&A...385..337T}, 
which solves the compressible equations of ideal magneto-hydrodynamics (MHD) in presence of gravity by employing a MUSCL-Hancock scheme with constrained transport on a finite volume Cartesian mesh \citep{2006JCoPh.218...44T, 2006A&A...457..371F}. 
We used a HLLC and a HLLD approximate Riemann solver for purely hydrodynamical and MHD simulations, respectively. The advantage of this type of solvers is that they resolve contact discontinuities explicitly, which is particularly important in highly subsonic flows to reduce numerical diffusivity.
RAMSES makes intensive use of the Message Passing Interface (MPI) library and it can therefore be used on massively parallel architectures. It is designed for high-resolution numerical simulations of a wide range of astrophysical problems, such as cosmology, galaxy and structure formation and evolution, and star formation.

However, the features of convective flows in stars pose a number of supplementary challenges that numerical schemes need to overcome. The pressure gradients in stellar interiors are balanced by gravity so that the whole system can attain a stationary configuration, that is hydrostatic equilibrium. Turbulent thermal convection is then often characterized by small-amplitude perturbations close to the equilibrium profile. As a consequence, we sought a numerical scheme that is able to preserve the hydrostatic equilibrium profiles in nonperturbed setups. Moreover, the typical velocities arising from turbulent convection are expected to be highly subsonic, which can be troublesome for time-explicit schemes.
Therefore, we had to adapt the numerical scheme of RAMSES to deal with highly subsonic, close to hydrostatic equilibrium turbulent flows, and we achieved that by implementing a well-balanced method. 
%
%
\subsection{Well-balanced scheme} 
\label{subsec:well-balances_scheme}
A well-balanced scheme numerically ensures the dynamical preservation of an equilibrium state by implicitly including it in the underlying discrete set of equations \citep[][]{1996SIAMJ.33...1G}. 
Such methods have been mainly developed in the context of shallow-water simulations \citep[see, e.g.,][]{2007JCoPh.226...29N}, 
but they are also becoming popular to tackle astrophysical problems \citep[see, e.g.,][]{2012JCoPh.231..919F, 2016A&A...587A..94K, 2019CiCP..26..1V, 2021A&A...652A..53E}. 

For the sake of simplicity, let us consider the one-dimensional Euler-Poisson set of equations, 
\begin{align}
     & \frac{\partial \rho}{\partial t} + \frac{\partial \left( \rho u \right)}{\partial x} = 0 \,, \nonumber\\
     & \frac{\partial u}{\partial t} + u \frac{\partial u}{\partial x} + \frac{1}{\rho} \frac{\partial p}{\partial x} = - \frac{\partial \phi}{\partial x} \,, \nonumber\\
    & \frac{\partial p}{\partial t} + u \frac{\partial p}{\partial x} + \gamma p \frac{\partial u}{\partial x} = 0 \,, \label{eq:euler_poisson_1D}
\end{align}
where $\rho$ is the density, $u$ is the velocity, $p$ is the thermal pressure, $\phi$ is the gravitational potential, and $\gamma$ is the adiabatic index of the ideal gas equation of state. The hydrostatic equilibrium equation then reads,
\begin{equation}
    \frac{\partial p}{\partial x} = - \rho \frac{\partial \phi}{\partial x} \,. \label{eq:hydro_eq}
\end{equation}
Standard finite volume methods struggle to accurately describe convective flows in stratified fluids because a discrete version of Eq.\,(\ref{eq:hydro_eq}) may not be preserved.
Hence, truncation errors introduced in the supposedly equilibrium states are interpreted as propagating waves by the numerical solvers.
These spurious waves can conceal the small perturbations of interest, disrupt the equilibrium states, and even undermine the numerical stability of the simulation. 
The numerical resolution required to prevent the truncation errors from spoiling the results would make the simulations easily impractical.

A possible solution is to impose the hydrostatic equilibrium given by Eq.\,(\ref{eq:hydro_eq}) directly in the set of equations governing the dynamics, that is Eq.\,(\ref{eq:euler_poisson_1D}). We separate the primitive variables of the problem into equilibrium, stationary states and dynamical perturbations,
\begin{equation}
    \rho = \tilde{\rho}+ \rho^{\prime} \,,~ p = \tilde{p} + p^{\prime} \,,~ u = u^{\prime}\,, \label{eq:hydro_var_separation} 
\end{equation}
where the equilibrium states of density and pressure, $\tilde{\rho}$ and $\tilde{p}$, exactly satisfy the hydrostatic equilibrium, Eq.\,(\ref{eq:hydro_eq}), while the velocity field $u$ is made up of perturbations only since $\tilde{u}=0$. With this separation at hand, we can rewrite the Euler-Poisson equations as,
\begin{align}
     & \frac{\partial \rho^{\prime}}{\partial t} + \frac{\partial \left( \rho u \right)}{\partial x} = 0 \,, \nonumber\\
     & \frac{\partial u}{\partial t} + u \frac{\partial u}{\partial x} + \frac{1}{\rho} \frac{\partial p^{\prime}}{\partial x} = - \frac{\rho^{\prime}}{\rho} \frac{\partial \phi}{\partial x} \,, \nonumber\\
    & \frac{\partial p^{\prime}}{\partial t} + u \frac{\partial p}{\partial x} + \gamma p \frac{\partial u}{\partial x} = 0 \,, \label{eq:euler_poisson_1D_well_balanced}
\end{align}
where the hydrostatic equilibrium in stationary regimes, that is where $\rho^{\prime}, p^{\prime}, u = 0$, is explicitly preserved.

Well-balanced schemes have already been used in the context of stellar convection by \citet{2019SciA....5.2307H}, \citet{2020A&A...641A..18H}, \citet{2021A&A...653A..55H}, and \citet{2021A&A...652A..53E}, for example. 
For this work, we modified both the hydrodynamical and MHD numerical solvers of RAMSES to account for a well-balanced version of the evolution equations. Indeed, we extended the well-balanced methodology also to the equations of ideal MHD, where we assume the equilibrium state to be characterized by $\tilde{\boldsymbol{B}} = 0$. 
Hence, magnetic fields are treated as pure perturbations, $\boldsymbol{B} = \boldsymbol{B}^{\prime}$,
and they do not modify the initial hydrostatic equilibrium, Eq.\,(\ref{eq:hydro_eq}). 
Moreover, we solve for the conservation equation of specific entropy $s = p/\rho^{\gamma}$ instead of total energy, that is,
\begin{equation}
    \frac{\partial s}{\partial t} + u\frac{\partial s  }{\partial x} = 0\,. \label{eq:cons_entropy}
\end{equation}
As a consequence, the total energy of the system may not be exactly conserved. On the other hand, we can accurately follow the dynamics of entropy perturbations which stem from the convective flows in adiabatic stratifications. The validity of this approach is guaranteed by the highly subsonic nature of typical stellar convective flows, where shocks are essentially absent \citep{2015ApJ...798...49W}. 

%
%
\subsection{Simulation setup} 
\label{subsec:setup}
\begin{table}
    \centering 
    \caption{Initial conditions of the He shell flash convective region model.}
    \def\arraystretch{1.3}
    \begin{tabular}{c c c c}
        \hline
        \hline 
          & Bottom domain         & Convection zone       & Top domain             \\ \hline 
        $z$         & $0.00$                & $1.64 \times 10^8$    & $7.77 \times 10^8$       \\
        $\rho$      & $6.13 \times 10^5$    & $1.17 \times 10^4$    & $9.15 \times 10^2$       \\
        $p$         & $5.14 \times 10^{20}$ & $1.77 \times 10^{20}$ & $2.51 \times 10^{18}$    \\
        $s$         & $9.81 \times 10^8$    & $1.19 \times 10^9$    & $1.19 \times 10^9$ \\
        $T$         & $1.40 \times 10^8$    & $2.47 \times 10^8$    & $4.53 \times 10^7$ \\
        $\Gamma$    & $1.20$                & $1.67$                & $1.01$                    \\ \hline 
    \end{tabular}
    \label{tab:IC}
    \tablefoot{The values of height $z$, density $\rho$, pressure $p$, specific entropy $s$, temperature $T$, and polytropic index $\Gamma$ refer to the bottom of the respective domains and are given in cgs units.} 
\end{table}
\begin{figure}
	\centering
	\resizebox{\hsize}{!}{\includegraphics{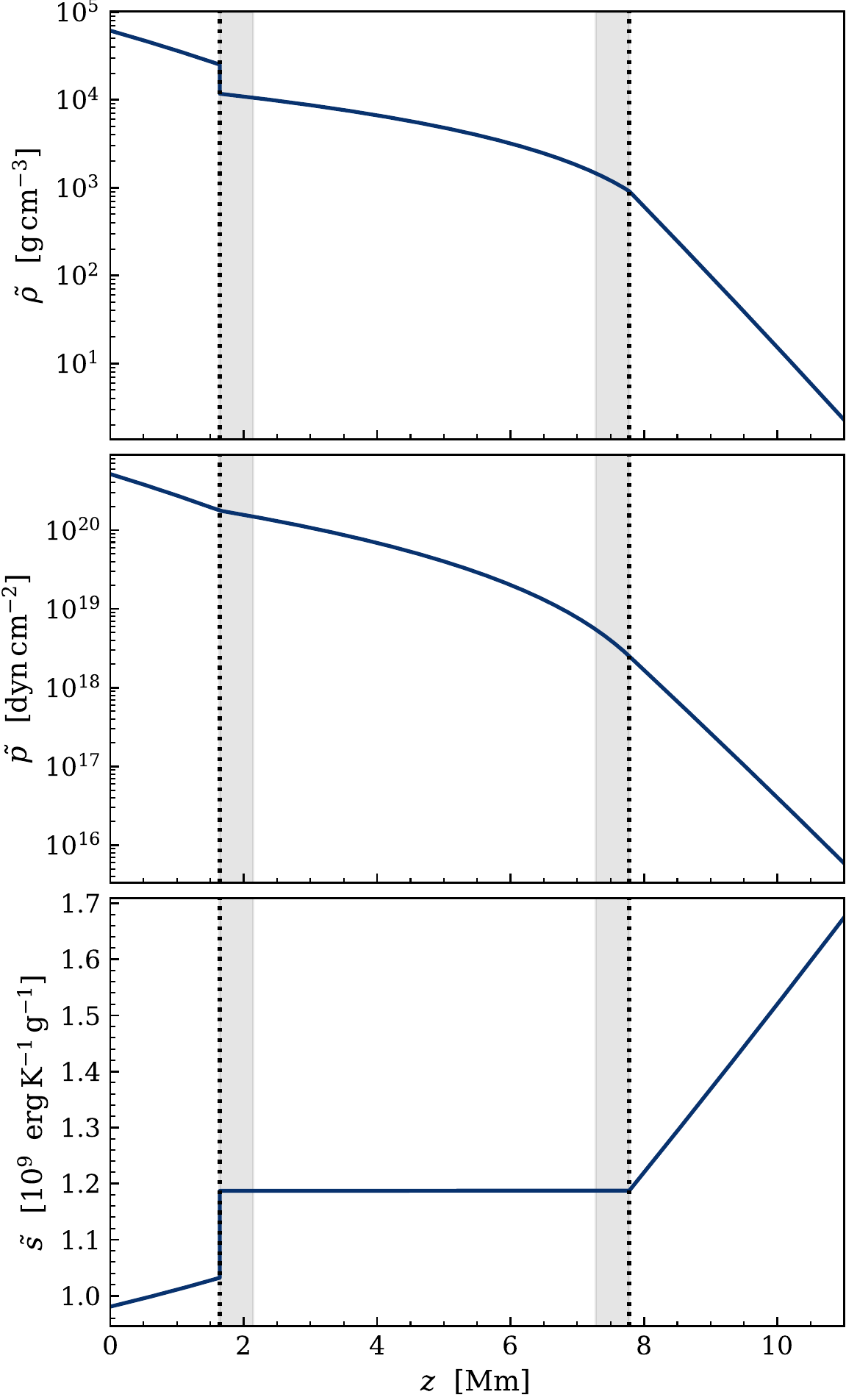}}
	\caption{Initial and equilibrium profiles of density $\tilde{\rho}$, pressure $\tilde{p}$, and specific entropy $\tilde{s}$. Black dotted lines delimit the convection zone, while the gray areas represent the heating (bottom) and cooling (top) layers.}
	\label{fig:IC}
\end{figure}

We aim to test our code on a simple but realistic, plane-parallel, stellar convective region. 
We chose to follow the initial setup presented by \citet{2006ApJ...642.1057H}, 
where an intershell of a typical low-mass AGB star near the He shell flash luminosity peak is modeled. 
The simulation box covers a cubic domain of dimensions $x,y,z \in \left[ 0,L \right]$, where $L = 11.0 \, {\rm Mm}$. 
This box is located at a radius of $7.51 \, {\rm Mm}$ of a stellar model, with main-sequence initial mass of $2\, M_{\odot}$ and metallicity $Z = 0.01$, which undergoes its second-to-last thermal pulse \citep[][]{2004ApJ...613L..73H}. 

The box is divided vertically in three domains which approximate the stellar stratification: a bottom stable region that extends up to $z = 1.64\,{\rm Mm}$, a convection zone in the middle, and an upper stable region from $z = 7.77\,{\rm Mm}$ to the top boundary. 
We initialized each domain with a polytropic stratification in hydrostatic equilibrium. 
The properties of the stratification at the bottom of each domain, as well as the respective polytropic indices, are given in Table\,\ref{tab:IC}. The corresponding equilibrium profiles for density, pressure, and specific entropy are shown in Fig.\,\ref{fig:IC}. We also assumed constant gravity directed downward along the vertical axis ($z$) with value $g = 10^{7.7}\,{\rm cm}\,{\rm s}^{-2}$ and a mono-atomic ideal gas with adiabatic index $\gamma = 5/3$ and equation of state
\begin{equation}
    p = \left(\gamma-1\right)\rho\left(\mathrm{e} - \frac{1}{2}u^2\right)\,, \label{eq:eos}
\end{equation}
where $\mathrm{e}$ is the specific total energy. 
For more information about the stellar model, the reader can refer to \citet{2006ApJ...642.1057H}. 

If the simulations are initialized with the equilibrium profiles of Fig.\,\ref{fig:IC}, the code is able to preserve the initial state indefinitely up to machine precision thanks to the implemented well-balanced scheme.
Hence, to set off convection, we introduced random density perturbations $\delta \rho$ of the order $\delta \rho / \tilde{\rho} \sim 10^{-2}$ in every grid cell within $1\,{\rm Mm}$ from the bottom of the convection zone. 

Turbulent convective motions are then sustained by nuclear reactions heating at the bottom and radiative cooling at the top, which we modeled by adding and removing energy in two bands of thickness $\Delta z = 0.5\,{\rm Mm}$ at the bottom and top, respectively, as shown in gray in Fig.\,\ref{fig:IC}. We note that  \citet[][]{2006ApJ...642.1057H} 
did not include radiative cooling, therefore our setup resembles that of surface convection.
We assumed constant and equal volumetric heating and cooling rates given by $\dot{e} = \dot{\epsilon}_{0}\,\rho_0$, where $\dot{\epsilon}_0 = 2 \times 10^{10}\,{\rm erg}\,{\rm g}^{-1}\,{\rm s}^{-1}$ and $\rho_0$ is the density at the base of the convection zone given in Table\,\ref{tab:IC}. Consequently, the energy flux at the bottom of the convection zone is $F = \dot{e} \Delta z = 1.17 \times 10^{22}\,{\rm erg}\,{\rm s}^{-1}\,{\rm cm}^{-2}$. This value accounts for the integrated amount of energy released according to the stellar model and corresponds to a stellar luminosity of $L = 3.21 \times 10^7\,L_{\odot}$. We note that the stellar luminosity is often boosted by several orders of magnitude in anelastic numerical simulations to balance the effect of high artificial viscosities \citep[see, e.g.,][]{2013ApJ...772...21R}. 
Since we are using a fully compressible approach, we can stick to more realistic values of $L$. 

We ran hydrodynamical (\texttt{HD}) and magneto-hydrodynamical (\texttt{MHD}) numerical simulations of the convective envelope described above. In both cases, we used the well-balanced MHD solver of RAMSES to allow for a direct comparison between the two. 
The hydrodynamical simulations are initialized with $\boldsymbol{B} = (0,0,0)\,{\rm G}$, ensuring that the magnetic field strength will remain zero everywhere throughout the simulation. 
For the magneto-hydrodynamical cases, we set a constant, homogeneous, and horizontal (along the $x$ axis) magnetic field of strength $B_0 = 10^3\,{\rm G}$ in the convection zone alone. In the stable domains instead, $\boldsymbol{B} = (0,0,0)\,{\rm G}$. 
This configuration preserves exactly the constraint $\boldsymbol{\nabla} \cdot \boldsymbol{B} = 0$. Moreover, the initial magnetic field is weak enough for a turbulent small-scale dynamo in the convection zone to amplify it significantly, as the expected equipartition value of the magnetic field strength for this setup is $B_{\rm eq} \sim 10^8\,{\rm G}$. 
We adopted fixed boundary conditions in the vertical direction by forcing the hydrodynamical variables to follow the polytropic equilibrium profiles, while the magnetic fields are forced to be zero in the ghost cells. The lateral boundary conditions are periodic for all variables.
We ran both hydrodynamical and magneto-hydrodynamical simulations with resolutions $N = 64^3,\,128^3,\,256^3,$ and $512^3$.
We identify the different simulations by their type and resolution (e.g., \texttt{HD\_256}).
In this study, we did not use the AMR capabilities of RAMSES. Therefore the grid is equidistant in each direction with cell sizes ranging from $21 \, {\rm km}$ to $172 \, {\rm km}$. 

%
%
\section{Results and discussion} 
\label{sec:results}
In this section, we present and discuss the results of our simulations. We focus on the onset of convection, on the turbulent amplification of the magnetic energy, and on the general properties of the convective flows and magnetic fields. Finally, we investigate the propagation of pressure and internal gravity waves in the stable regions.   
%
%
\subsection{Onset of convection} 
\label{subsec:onset_of_convection}

\begin{figure*}
	\centering
	\resizebox{\hsize}{!}{\includegraphics{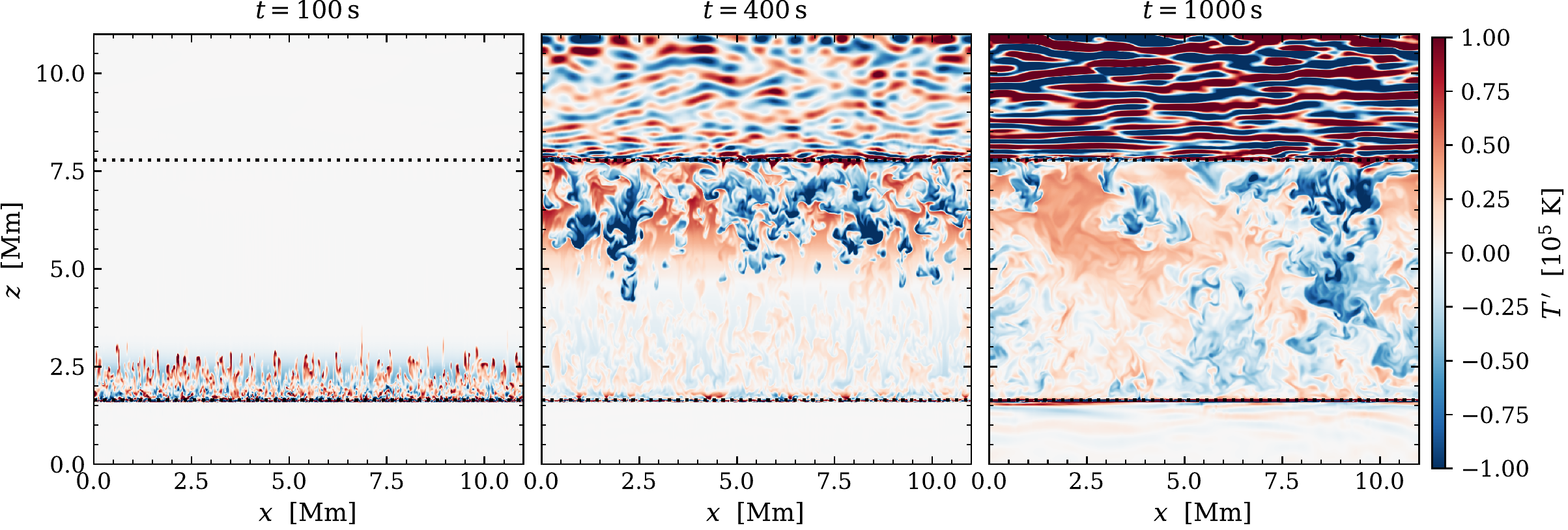}}
	\caption{Temperature fluctuations $T^{\prime}$ in a vertical section of the \texttt{HD\_512} simulation at $t = 100\,{\rm s}$ (\textit{left}), $t = 400\,{\rm s}$ (\textit{middle}), and $t = 1\,000\,{\rm s}$ (\textit{right}). The vertical section is taken at $y = 3.43\,{\rm Mm}$. Black dotted lines indicate the boundaries of the convection zone.}
	\label{fig:convection_visualization}
\end{figure*}
\begin{figure}
	\centering
	\resizebox{\hsize}{!}{\includegraphics{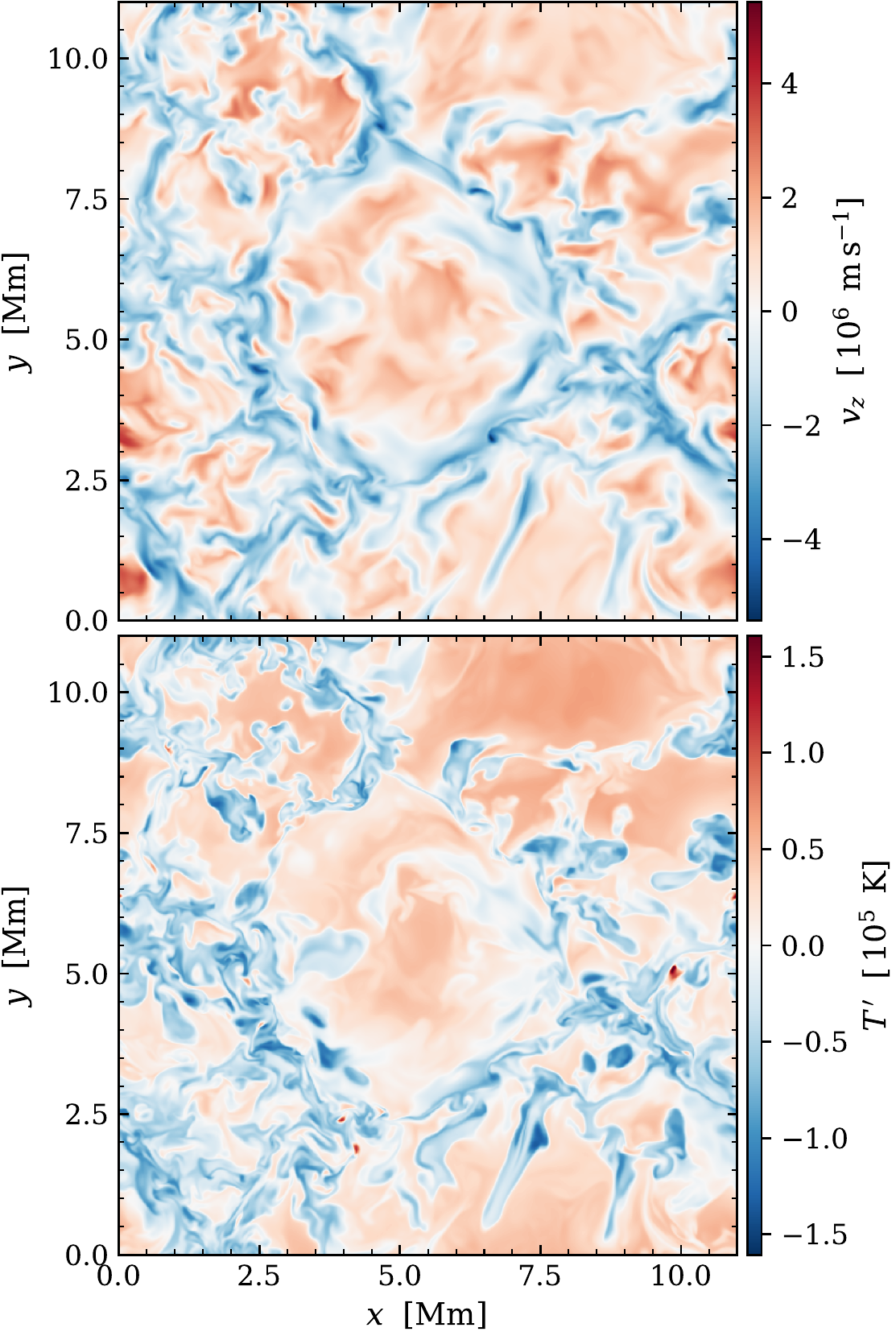}}
	\caption{Horizontal sections of the vertical velocity $v_z$ (top) and temperature fluctuation $T^{\prime}$ (bottom) in the upper layers of the convection zone ($z = 7.4\,{\rm Mm}$). The snapshot is taken from the \texttt{HD\_512} simulation at $t = 1\,000\,{\rm s}$.}
	\label{fig:convection_granules}
\end{figure}
\begin{figure}
	\centering
	\resizebox{\hsize}{!}{\includegraphics{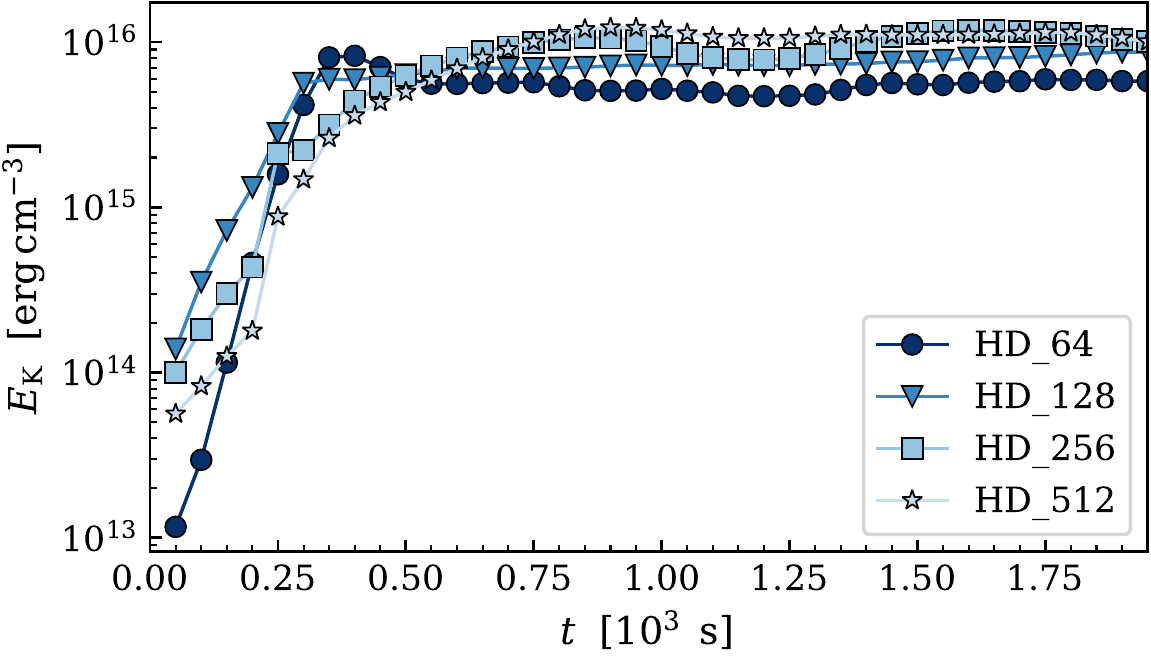}}
	\caption{Time evolution of the mean turbulent kinetic energy density $E_{\rm K}$ in the convection zone for the four hydrodynamical (\texttt{HD}) simulations.}
	\label{fig:convection_energy}
\end{figure}
\begin{figure}
	\centering
	\resizebox{\hsize}{!}{\includegraphics{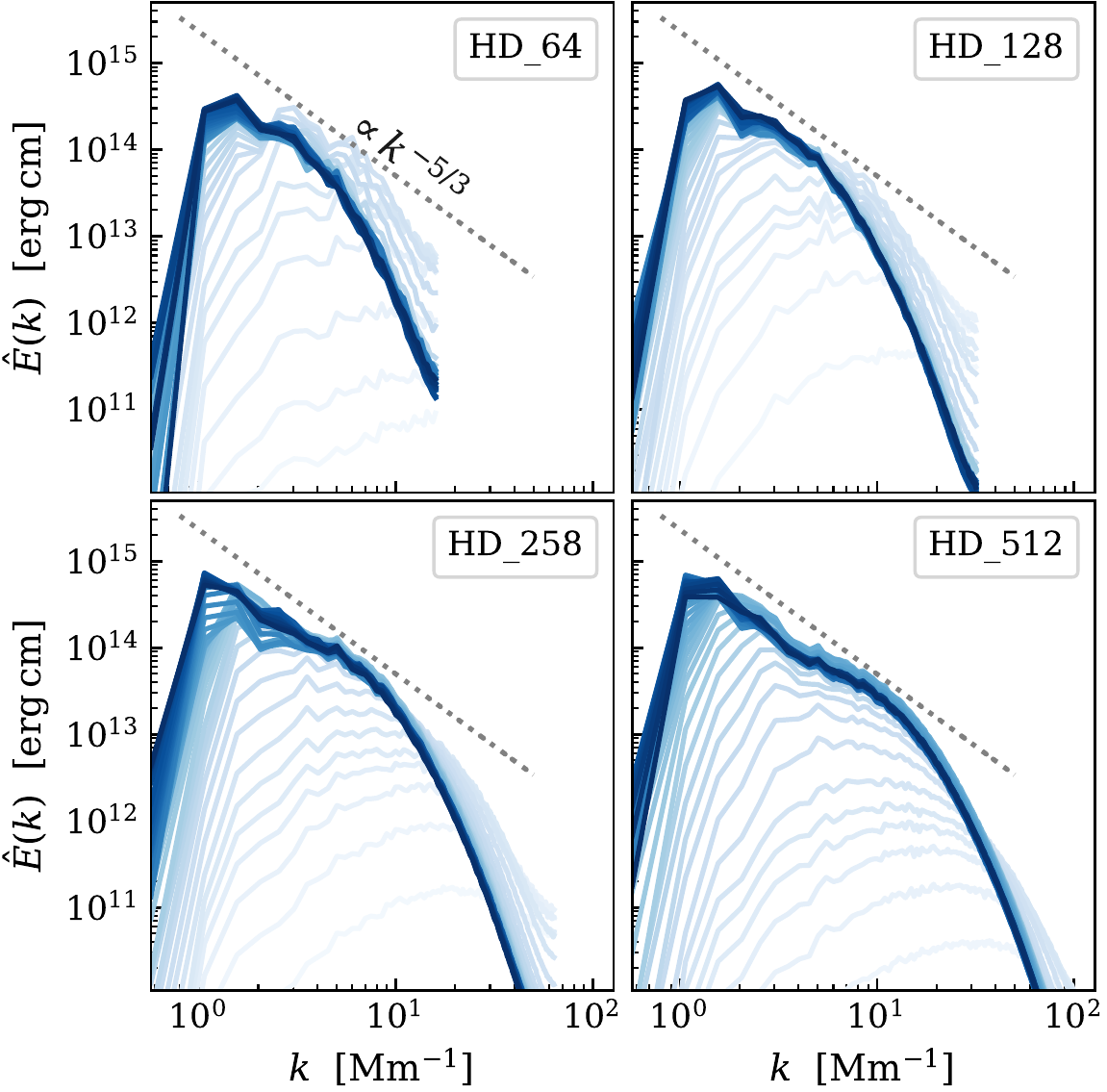}}
	\caption{Time evolution of the turbulent kinetic energy power spectra $\hat{E}_{\rm K}$ for the four \texttt{HD} simulations. The time evolution is represented by the color grading: light blues correspond to early times, while the darkest shade corresponds to $t = 2\,000\,{\rm s}$. The interval between each color-shade is $50\,{\rm s}$.}
	\label{fig:convection_PS_evolution}
\end{figure}
\begin{figure}
	\centering
	\resizebox{\hsize}{!}{\includegraphics{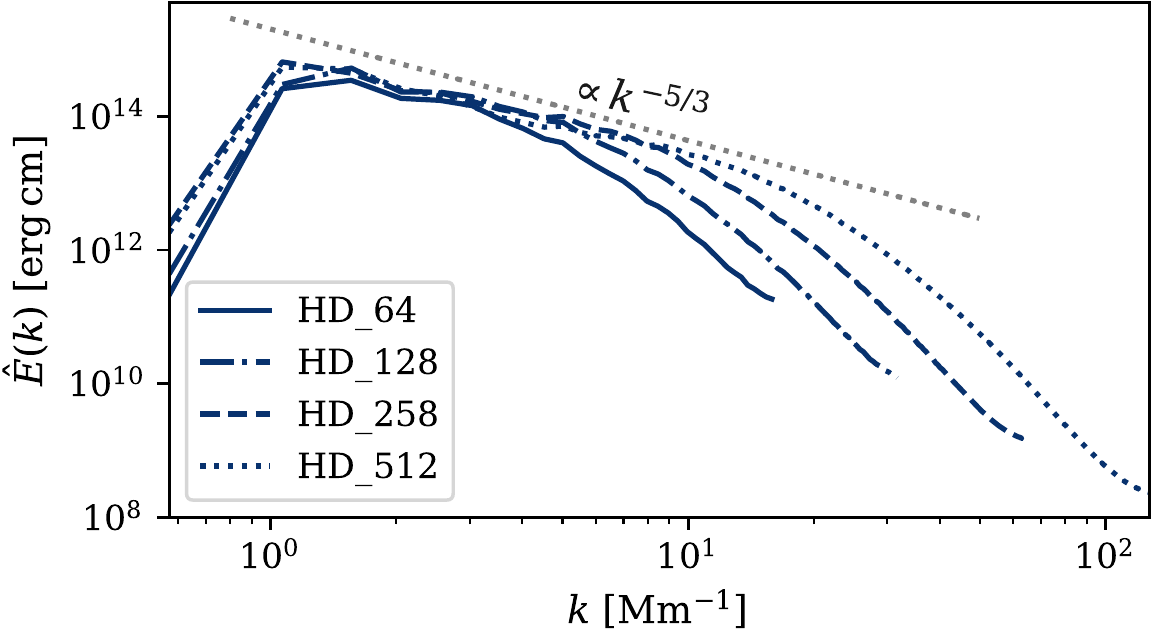}}
	\caption{Comparison of the kinetic energy power spectra $\hat{E}_{\rm K}$ between the four \texttt{HD} simulations in the quasi-steady state regime. The profiles are obtained by averaging the power spectrum profiles over the last 10 snapshots of each simulation, that is between $t=1\,500\,{\rm s}$ and $t=2\,000\,{\rm s}$.}
	\label{fig:convection_PS_comparison}
\end{figure}

We start by analyzing the onset of convection in the hydrodynamical simulations, \texttt{HD\_64}, \texttt{HD\_128}, \texttt{HD\_256}, and \texttt{HD\_512}, which we ran for a total of $2\,000\,{\rm s}$ physical time. We define the convective turnover time scale as,
\begin{equation}
    \tau_{\rm conv} = \frac{2 L_{\rm CZ}}{v_{\rm conv}} \sim 1\,000\,{\rm s}\,, \label{eq:convective_time} 
\end{equation}
where $L_{\rm CZ} \sim 6\,{\rm Mm}$ is the height of the convection zone and $v_{\rm conv}\sim 1.2\times10^6\,{\rm cm}\,{\rm s}^{-1}$ is the characteristic convective velocity (see Sect.\,\ref{subsec:velocity_and_magnetic_fields}). Hence, our hydrodynamical simulations cover $\sim 2$ convective turnover times.

Figure \ref{fig:convection_visualization} shows three vertical sections of the temperature fluctuations, defined as $T^{\prime} = T - \tilde{T}$, at times $t = 100,\, 400,$ and $1\,000\,{\rm s}$, respectively, for the $\texttt{HD\_512}$ simulation. Small-scale thermal convective instabilities developing from the initially perturbed cells can be seen at time $t = 100\,{\rm s}$. 
These finger-like structures rise through the convection zone at slightly higher temperatures than the equilibrium values, on the order of $\sim 0.1\,\%$. 
The growth of these instabilities also excite small amplitude pressure waves that sweep the convection zone and reach the top stable layer. These waves induce tiny perturbations in the equilibrium profiles, that are not visible in the left panel of Fig.\,\ref{fig:convection_visualization} but seed a second cascade of thermal convective 
instabilities. This cascade is driven by the external radiative cooling layer, as we can see in the middle panel of the same figure, and it is characterized by descending turbulent filaments with negative temperature fluctuations around the equilibrium profile \citep[see, e.g.,][]{2013ApJ...769....1V}. 

After $t \sim 1\,000\,{\rm s}$, we attain a fully developed convection zone, which is maintained thereafter by the balance between external cooling and heating. Indeed, when the rising plasma heated at the bottom reaches the top of the convection zone, it is cooled down and initiates a turbulent downflow (see right panel of Fig.\,\ref{fig:convection_visualization}). The large-scale structure of the flow is therefore granular-like, with large and hot slowly uprising plumes surrounded by narrow and cold filamentary downflows. 

In Fig.\,\ref{fig:convection_granules} we show a horizontal section of the vertical velocity $v_z$ and of the temperature fluctuations $T^{\prime}$ near the top of the convection zone at $t = 1\,000\,{\rm s}$. A granulation pattern, typical of surface convection \citep[see, e.g.,][]{1998ApJ...499..914S}, 
is clearly observable 
with the central granule having an approximate diameter of $\sim 5\,{\rm Mm}$. 
The large-scale convective motions buffet the stable plasma in the bottom and top stable layers, exciting there different oscillation modes that are visible as horizontally extending temperature fluctuations in the central and right panels of Fig.\,\ref{fig:convection_visualization}.
We address the nature of these oscillations in Sect.\,(\ref{subsec:waves}).

It is important to notice that the dynamical fluctuations characterizing the convective region are small if compared to the equilibrium profiles values. 
Figures \ref{fig:convection_visualization} and \ref{fig:convection_granules} show temperature fluctuations with typical values around 3 orders of magnitude smaller than the equilibrium values in the convection zone. 
Similar amplitudes are also valid for the density and pressure perturbations, while the typical vertical velocities at the top of the convection zone are 2 orders of magnitude smaller than the local sound speed $c_{\rm s}$. 
Therefore, the well-balanced scheme implemented in the code and presented in Sect.\,(\ref{subsec:well-balances_scheme}) is crucial to correctly capture the dynamics of this problem.

Figure \ref{fig:convection_energy} shows, for the four hydrodynamical simulations, the time evolution of the mean turbulent kinetic energy density, $E_{\rm K} = \frac{1}{2}\rho v_{\rm rms}^2$, where $v_{\rm rms}$ is the root-mean-squared (rms) turbulent three-dimensional velocity field strength. Since the two stable layers present no significant turbulent motions, we restrict the computation of $E_{\rm K}$ to the convection zone alone. 
The onset of convection is represented by an initial exponential growth, which lasts between $\sim 250\,{\rm s}$ and $\sim 1\,000\,{\rm s}$, depending on the resolution. 
Then, a quasi-steady state is reached, characterized by a quasi-constant mean turbulent kinetic energy density. 
The runs \texttt{HD\_128}, \texttt{HD\_256}, and \texttt{HD\_512} reach a mean turbulent kinetic energy density of around $10^{16}\,{\rm erg}\,{\rm cm}^{-3}$, while the low resolution one, \texttt{HD\_64}, stabilizes around a slightly lower value. 
We notice long lived fluctuations in the quasi-steady state mean kinetic energy density that are due to the episodic and large-scale nature of the flow in the convection zone \citep[][]{2007ApJ...667..448M}. 

The growth rate of the mean kinetic energy density seems to decrease with increasing resolution: the low-resolution simulations ($\texttt{HD\_64}$ and $\texttt{HD\_128}$) show a fast and sharp evolution to a quasi-steady state, while a smooth and slower growth characterizes the high-resolution ones ($\texttt{HD\_256}$ and $\texttt{HD\_512}$). 
We can qualitatively explain this difference with the transition between initial instabilities and large-scale convective flows. In fact, the thermal convective 
instabilities at the origin of convection form at the smallest scales in the simulation, since the random perturbations in density are introduced at the grid-size level. On the other hand, the kinetic energy dominant structures are the large-scale ones, that is plumes and downflows. This results in an apparent slower growth since the smaller the initial scale of the turbulence, the longer it will take to form steady state, large-scale convection.

The transition process between small and large scales can also be observed in Fig.\,\ref{fig:convection_PS_evolution}, where we show the time evolution of the turbulent kinetic energy density power spectra, $\hat{E}_{\rm K}$, for the four hydrodynamical simulations. 
We notice that in the first stages of the simulations (light blue), the peaks of the various power spectra are found at large wavenumber $k$, that is at small scales in physical space. As the simulations evolve, the total kinetic energy densities grow and the peaks of the power spectra shifts towards small wavenumbers.

When the quasi-steady state is reached (dark blue profiles), the energy cascade follows a Kolmogorov power law with index $-5/3$ from the steady state energy peak up to the dissipation scale in all panels of Fig.\,\ref{fig:convection_PS_evolution}. 
The same can be observed in Fig.\,\ref{fig:convection_PS_comparison}, where we compare the average kinetic energy power spectra during the quasi-steady state phase between the different runs.
The better the spatial resolution, the larger the range where the Kolmogorov cascade is well reproduced. The quasi-steady state energy peak instead is found around $k\sim 1\,{\rm Mm}^{-1}$ for all resolutions. This value corresponds to the typical size of the large-scale uprising convective plumes observed in Fig.\,\ref{fig:convection_granules}, that is $l = 2 \pi/k \sim 6\,{\rm Mm}$.
%
%
\subsection{Small-scale dynamo} 
\label{subsec:dynamo}

\begin{figure}
	\centering
	\resizebox{\hsize}{!}{\includegraphics{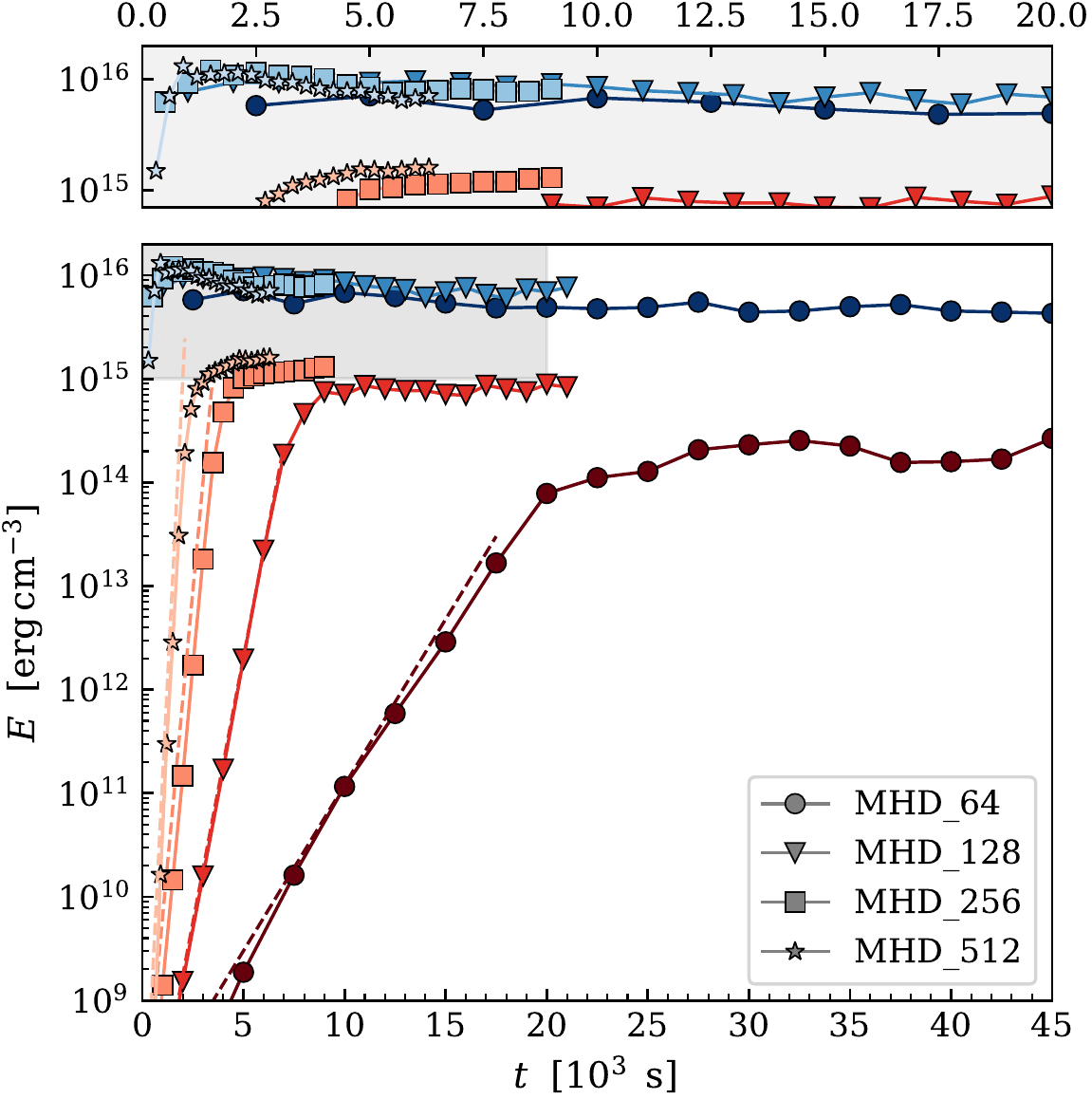}}
	\caption{Time evolution of the mean turbulent kinetic energy density $E_{\rm K}$ (blue) and of the mean magnetic energy density $E_{\rm M}$ (red) in the convection zone for the four magneto-hydrodynamical (\texttt{MHD}) simulations. Dashed lines represent exponential fits obtained with the kinematic growth rates $\gamma_{\rm K}$ listed in Table\,\ref{tab:equipartition_B}. The top panel is a zoom-in on the first $20\,000\,{\rm s}$ of the turbulent kinetic energy evolution.}
	\label{fig:dynamo_energy}
\end{figure}
\begin{figure}
	\centering
	\resizebox{\hsize}{!}{\includegraphics{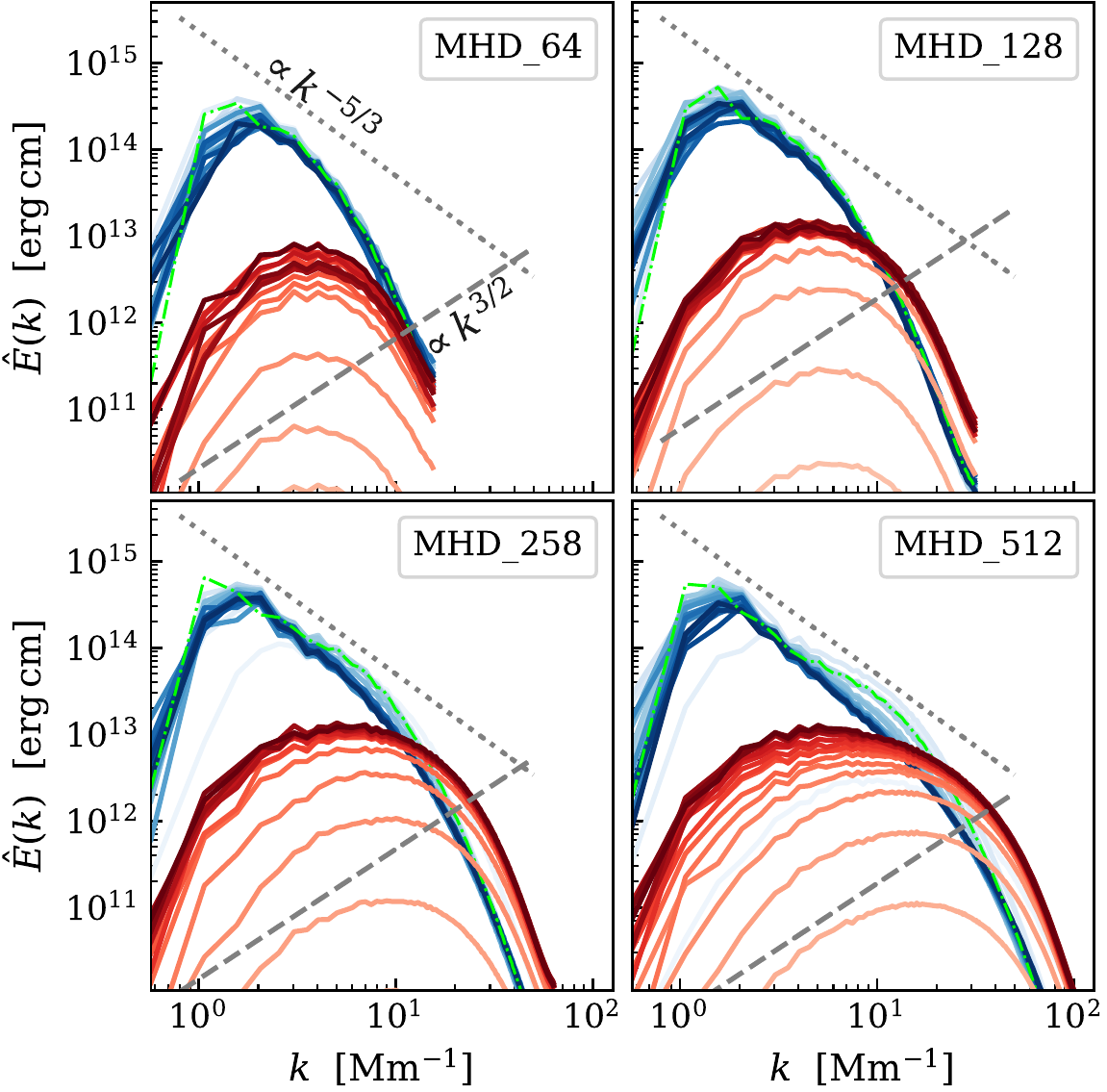}}
	\caption{Time evolution of the power spectra of the turbulent kinetic energy $\hat{E}_{\rm K}$ (blue) and magnetic energy $\hat{E}_{\rm M}$ (red) for the four magneto-hydrodynamical (\texttt{MHD}) simulations. The time evolution is represented by the color grading: light colors correspond to early times, while the darkest shade corresponds to the last snapshot of the simulation. The time interval between each shade is $5\,000\,{\rm s}$ for $\texttt{MHD\_64}$, $1\,000\,{\rm s}$ for $\texttt{MHD\_128}$, $500\,{\rm s}$ for $\texttt{MHD\_256}$, and $300\,{\rm s}$ for $\texttt{MHD\_512}$. The average quasi-steady state kinetic energy power spectrum for the hydrodynamical simulation $\texttt{HD\_512}$ is shown in green.}
	\label{fig:dynamo_PS_evolution}
\end{figure}
\begin{figure}
	\centering
	\resizebox{\hsize}{!}{\includegraphics{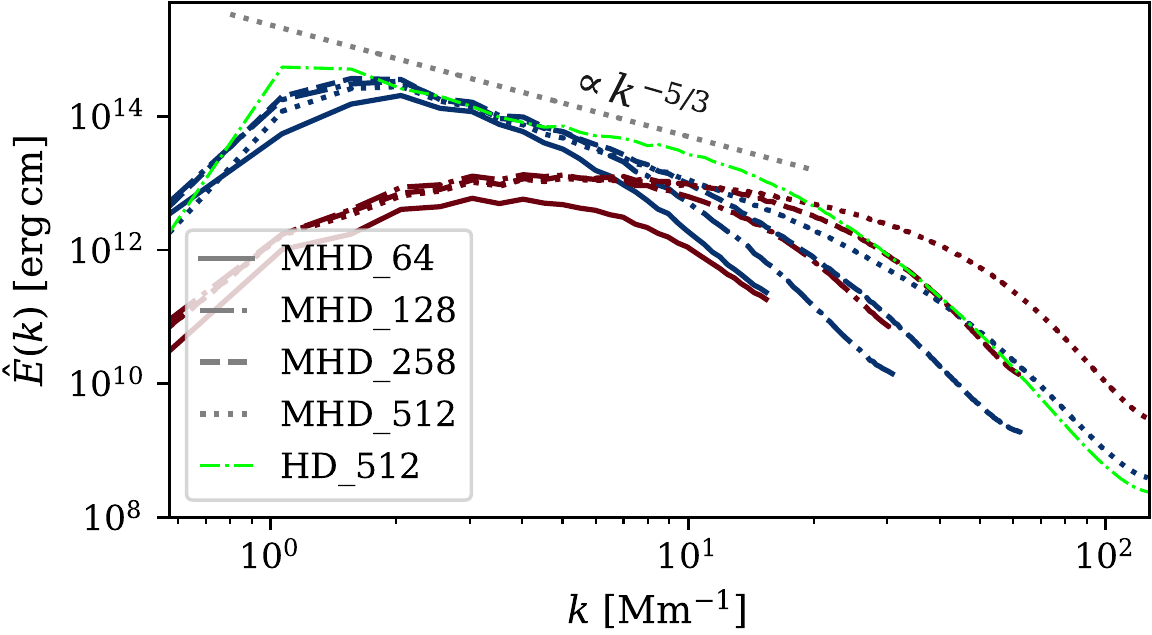}}
	\caption{Comparison between the power spectra of the turbulent kinetic energy (blue) and magnetic energy (red), $\hat{E}_{\rm K}$ and $\hat{E}_{\rm M}$, for the four magneto-hydrodynamical (\texttt{MHD}) simulations after saturation of the magnetic field. The average quasi-steady state kinetic energy power spectrum for the hydrodynamical simulation $\texttt{HD\_512}$ is shown in green.}
	\label{fig:dynamo_PS_comparison}
\end{figure}
\begin{figure*}
	\centering
	\resizebox{\hsize}{!}{\includegraphics{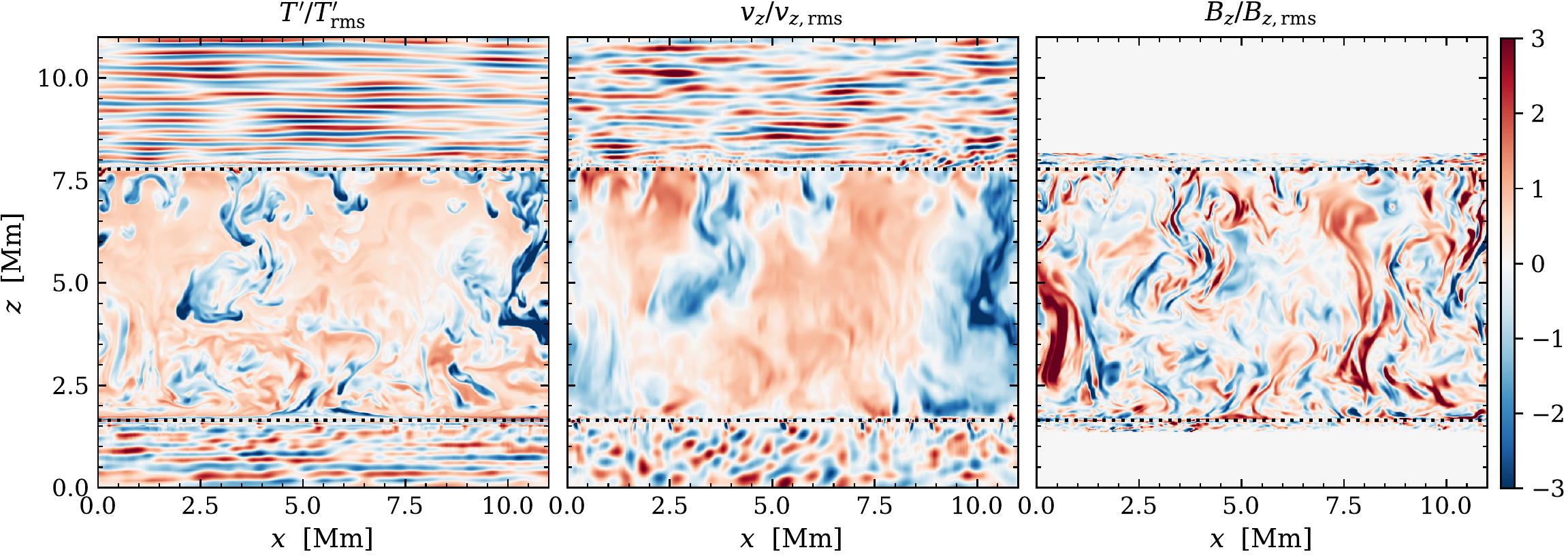}}
	\caption{Vertical sections of the temperature fluctuations $T^{\prime}$ (\textit{left}), vertical velocity $v_z$ (\textit{middle}), and vertical magnetic field $B_z$ (\textit{right}). To enhance the visibility of the perturbations, each quantity is scaled by its mean rms value at each height $z$. The sections are taken from the simulation $\texttt{MHD\_512}$ at $t=5\,000\,{\rm s}$ and $y=2.34\,{\rm Mm}$. Black dotted lines indicate the boundaries of the convection zone.}
	\label{fig:dynamo_visualization}
\end{figure*}
\begin{figure}
	\centering
	\resizebox{\hsize}{!}{\includegraphics{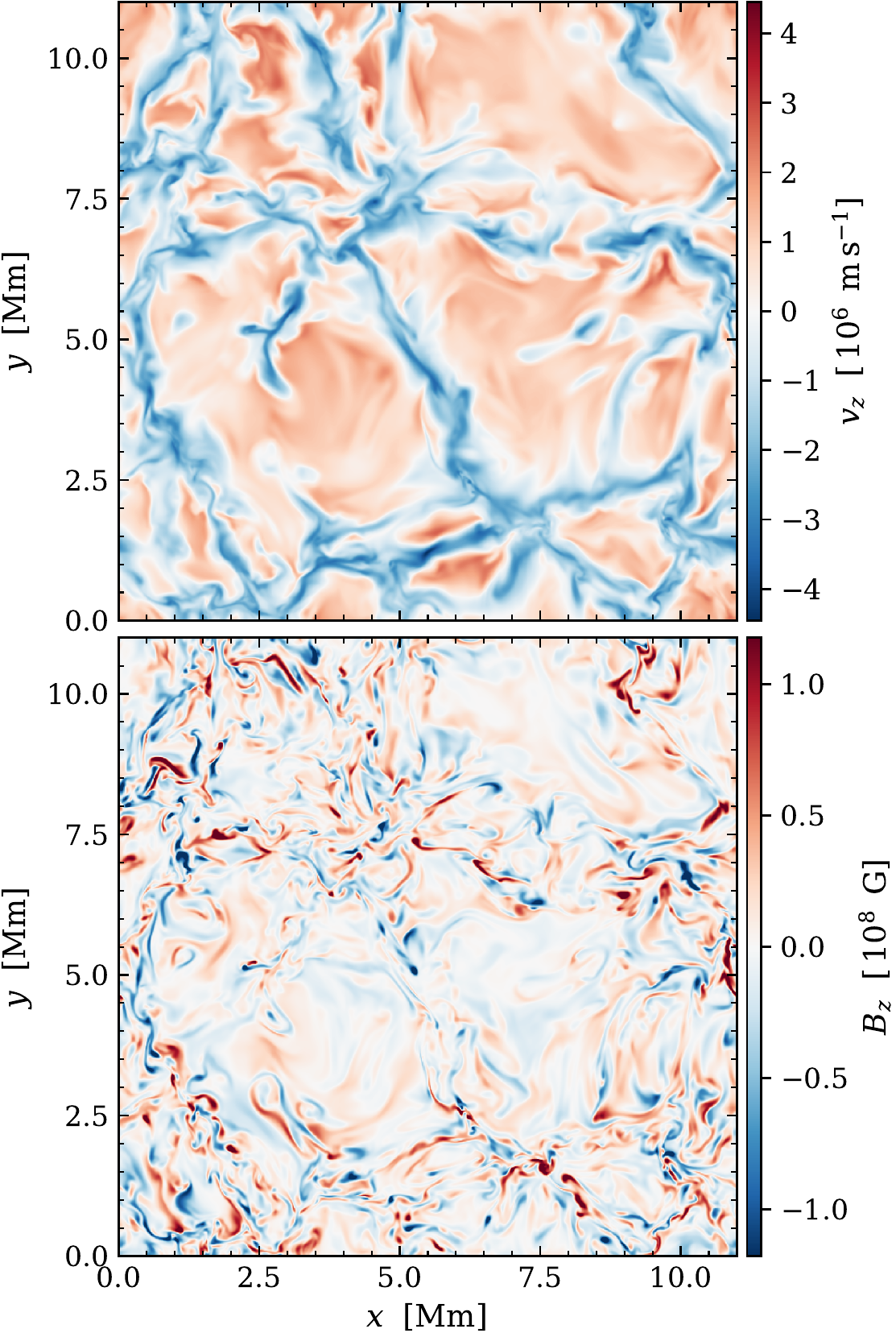}}
	\caption{Horizontal sections of the vertical velocity $v_z$ (\textit{top}) and vertical magnetic field $B_z$ (\textit{bottom}) near the top of the convection zone ($z=7.40\,{\rm Mm}$). The sections are taken from the $\texttt{MHD\_512}$ simulation at $t=5\,000\,{\rm s}$.}
	\label{fig:dynamo_granules}
\end{figure}

\begin{table}
    \centering 
    \caption{Kinematic growth rates $\gamma_{\rm K}$ and ratio between saturated rms magnetic field strength $B_{\rm rms}$ and equipartition value $B_{\rm eq}$ for the MHD simulations.}
    \def\arraystretch{1.3}
    \begin{tabular}{c | c c c c}
        \hline
        \hline 
         &\texttt{MHD\_64} & \texttt{MHD\_128} & \texttt{MHD\_256} & \texttt{MHD\_512} \\ \hline 
        $\gamma_{\rm K}~[10^{-3}~{\rm s}^{-1}]$ & $0.74$ & $2.35$ & $4.83$ & $8.87$  \\ \hline
        $B_{\rm rms}/B_{\rm eq}$ & $0.15$ & $0.25$ & $0.30$ & $0.36$ \\ \hline 
    \end{tabular}
    \label{tab:equipartition_B}
    \tablefoot{The growth rates are computed by fitting the magnetic energy densities over the first 8 snapshots of each simulation, while the ratios are averaged over the final snapshots of each simulation where the magnetic energy does not show any significant growth.} 
\end{table}

In this section we present the magneto-hydrodynamical simulations \texttt{MHD\_64}, \texttt{MHD\_128}, \texttt{MHD\_256}, and \texttt{MHD\_512}. 

Figure \ref{fig:dynamo_energy} shows the time evolution of the mean turbulent kinetic energy density $E_{\rm K}$ in blue, and of the mean magnetic energy density $E_{\rm M} = B^2/8\pi$ in red, for the four simulations.
Both energy densities are computed in the convection zone alone. The length of the simulations decreases with resolution because of the increase of computational cost. 
Nevertheless, we ran each simulation long enough for the magnetic field to be amplified and attain saturation. Indeed, while convection develops and reaches a quasi-steady state in a much shorter time scale, we observe in all four simulations a slower kinematic and a saturation phase: first the magnetic energy grows exponentially as $E_{\rm M} \sim \exp{(\gamma_{\rm K} t)}$, where $\gamma_{\rm K}$ is the kinematic growth rate, and then it reaches a quasi-constant value.

The different slopes show how the exponential growth depends on resolution, which is compatible with the action of a small-scale dynamo. In the first row of Table\,\ref{tab:equipartition_B} we show the kinematic growth rates $\gamma_{\rm K}$ of the different simulations. We find that the kinematic growth rate scales with resolution as $\gamma_{\rm K} \sim \Delta x^{-1.2}$ if we include all four simulations. However, the low-resolution simulation (\texttt{MHD\_64}) is not capturing all the dynamical scales of convection, as we have seen in Sect.\,\ref{subsec:onset_of_convection}. If we consider only runs \texttt{MHD\_128}, \texttt{MHD\_256}, and \texttt{MHD\_512}, the growth rate is $\gamma_{\rm K} \sim \Delta x^{-1.3}$. This result, which is consistent with what \citet{2010ApJ...714.1606P}, \citet{2014ApJ...789..132R}, and \citet{2022A&A...660A.115R} 
found for small-scale dynamo simulations of the quiet Sun magnetism, is very puzzling and in contrast with Kazantsev's dynamo theory prediction of $\gamma_{\rm K} \sim \Delta x^{-2/3}$.

The kinematic phase comes to an end when the magnetic field strength approaches the equipartition value, $B_{\rm eq} = \sqrt{4\pi \rho} v_{\rm rms}$. 
At this stage, the magnetic fields are strong enough to start back-reacting into the plasma dynamics by means of the Lorentz force. 
The mean magnetic energy eventually saturates and each simulation reaches a magneto-convective quasi-steady state. Typical magnetic field strengths in the convection zone at this stage are in the order of $B \sim 10^8\,{\rm G}$.
The ratios between the rms magnetic field strength $B_{\rm rms}$ and the equipartition magnetic field strength $B_{\rm eq}$ during the saturation phase are shown in Table\,\ref{tab:equipartition_B} and grow with increasing spatial resolution.

We show the time evolution of the turbulent kinetic and magnetic energy power spectra in Fig.\,\ref{fig:dynamo_PS_evolution}. 
We notice that the magnetic field strength is more efficiently amplified at small scales. 
During the kinematic phase (light red) in fact, the larger the spatial resolution, the more the peak of the magnetic energy power spectrum is shifted towards large $k$. 
In an ideal Kazantsev's dynamo, the magnetic energy power spectra should follow a power-law with index $3/2$ during the kinematic phase \citep[][]{2005PhR...417....1B}. 
In our case, the magnetic energy power spectra follow the Kazantsev's prediction only during the very early times of the kinematic phase. However, as they approach the saturation phase, they do not seem to follow such a profile.  

As the simulations evolve, the magnetic fields become dominant over the kinetic energy density at large $k$ for all the resolutions except the \texttt{MHD\_64}. The crossover scale is found around $k \sim 10\,{\rm Mm}^{-1}$ $(l\sim 2\,{\rm Mm})$.
The transition to super-equipartition of the magnetic fields at large $k$ coincides with a suppression of kinetic power at the same scales, which follows from the Lorentz-force feedback. Consequently, the kinetic power spectra profiles deviate from the Kolmogorov power law for $k \gtrsim 3\,{\rm Mm}^{-1}$ $(l \sim 0.5\,{\rm Mm})$.  We can appreciate the difference to the hydrodynamical simulation $\texttt{HD\_512}$ power spectrum which is plotted in green in Fig.\,\ref{fig:convection_PS_evolution}. Approaching the saturation phase (dark red), the magnetic energy spectrum peaks shift towards smaller wavenumbers and finally stabilizes around $k \sim 3\,{\rm Mm}^{-1}$, that corresponds to magnetic structures in physical space of size $l \sim 0.5\,{\rm Mm}$.

In Fig.\,\ref{fig:dynamo_PS_comparison} we show the kinetic and magnetic power spectra for the last snapshot of the four simulations. We notice that $\texttt{MHD\_128}$, $\texttt{MHD\_256}$, and $\texttt{MHD\_512}$ yield very similar results in the respective energy injection and inertial ranges, while the low-resolution simulation, $\texttt{MHD\_64}$, presents lower profiles for both kinetic and magnetic spectra at all scales. The kinetic energy power spectra of the MHD simulations clearly deviate from the hydrodynamical analog, $\texttt{HD\_512}$, shown in green. The suppression of kinetic energy is evident at around the injection scale ($k \sim 1 \, {\rm Mm}^{-1}$) and the crossover scale ($k \sim 10 \, {\rm Mm}^{-1}$). The power spectra evolution and steady state configuration for the high-resolution simulations are in qualitative agreement with the results obtained by \citet[][]{2014ApJ...789..132R} 
and \citet[][]{2015ApJ...803...42H} 
for small-scale dynamo studies in the solar convection zone. Therefore, the exponential growth of the magnetic field energy in our simulations is compatible with the action of a turbulent small-scale dynamo.

In Figure \ref{fig:dynamo_visualization} we show a vertical section of the temperature fluctuations $T^{\prime}$, vertical velocity $v_z$, and vertical magnetic field $B_z$ for the $\texttt{MHD\_512}$ simulation in the saturated phase. To enhance the visibility of the perturbations, we scaled each quantity by their mean rms value at each height $z$. However, in the stable regions where the magnetic field is very weak, spurious signals arise where $B_{z,{\rm rms}} \sim 0\,{\rm G}$. Therefore we masked out the regions where $B_{z,{\rm rms}} < 1\,{\rm G}$ in the third panel. For the same snapshot, we also plot a horizontal section of vertical velocity $v_z$ and vertical magnetic field $B_z$ close to the top the convection zone ($z=7.4\,{\rm Mm})$ in Fig.\,\ref{fig:dynamo_granules}.

The large-scale structures of the convective flow are the same ones we found in the hydrodynamical simulations: large and slowly ascending convective plumes surrounded by cold and narrow downflows, as we can see in the left and middle panels of Fig.\,\ref{fig:dynamo_visualization} and in the top panel of Fig.\,\ref{fig:dynamo_granules}. Moreover, as for the hydrodynamical simulations, we observe oscillating patterns in temperature and vertical velocity in the stable layers that hint to the presence of waves propagating there. 

In the right panel of Fig.\,\ref{fig:dynamo_visualization}, we see that the vertical magnetic field is characterized by small-scale filaments with mixed polarity. The typical thickness of these filaments is $l \sim 0.5\,{\rm Mm}$, which corresponds to the wavenumber ($k \sim 3\,{\rm Mm}^{-1}$) of the peaks in the magnetic energy power spectra seen in Fig.\,\ref{fig:convection_PS_evolution} and Fig.\,\ref{fig:convection_PS_comparison}.
Strong magnetic fields are found preferentially in intergranular downflows, as it can be seen in the bottom panel of Fig.\,\ref{fig:dynamo_granules} for the top of the convection zone. We also notice the presence of magnetic fields in thin layers above and below the convection zone in the right panel of Fig.\,\ref{fig:dynamo_visualization}. The magnetic field strength quickly decays to zero in these layers, but its presence demonstrates that the plasma is overshooting into the stable regions. Indeed, the frozen-in magnetic field generated in the turbulent convection zone could never be found in the stable layers without overshooting. 
%
\subsection{Velocity and magnetic fields properties} 
\label{subsec:velocity_and_magnetic_fields}

\begin{figure}
	\centering
	\resizebox{\hsize}{!}{\includegraphics{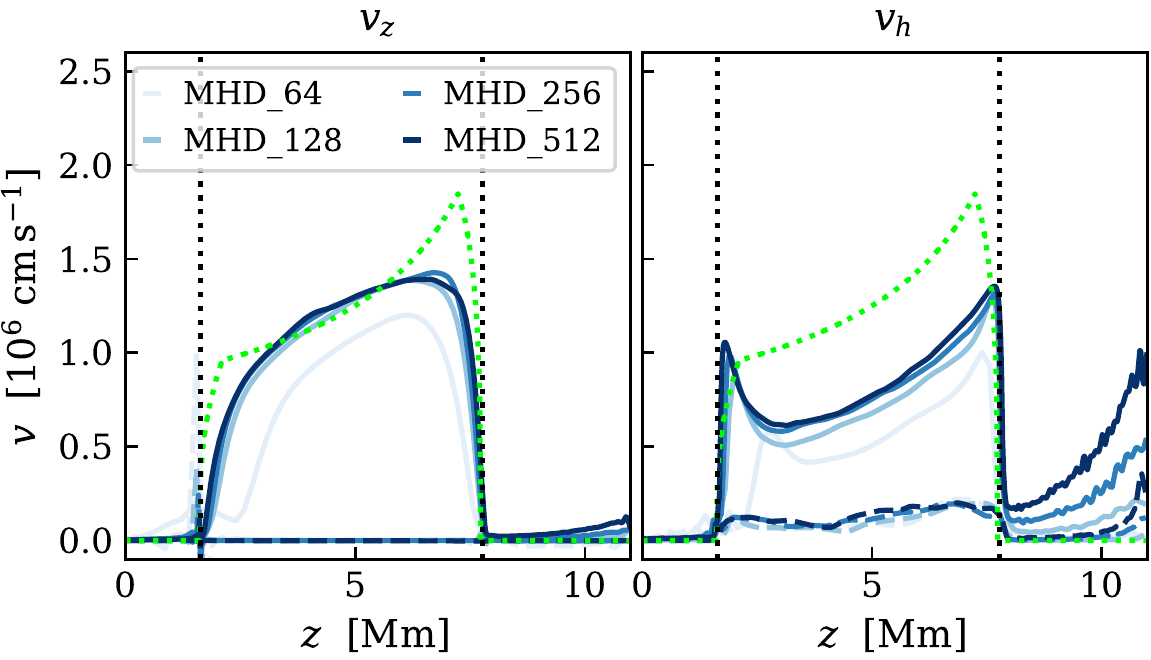}}
	\caption{Vertical profiles of the vertical (\textit{left}) and horizontal (\textit{right}) components of the velocity field for the four MHD simulations. The rms profiles are shown in continuous lines, while dashed lines represent the mean profiles. Dotted vertical lines denote the boundaries of the convection zone. The MLT prediction for the one-dimensional dispersion velocity given by Eq.\,(\ref{eq:MLT_sigma}) is shown in green.}
	\label{fig:properties_profiles_v}
\end{figure}
\begin{figure}
	\centering
	\resizebox{\hsize}{!}{\includegraphics{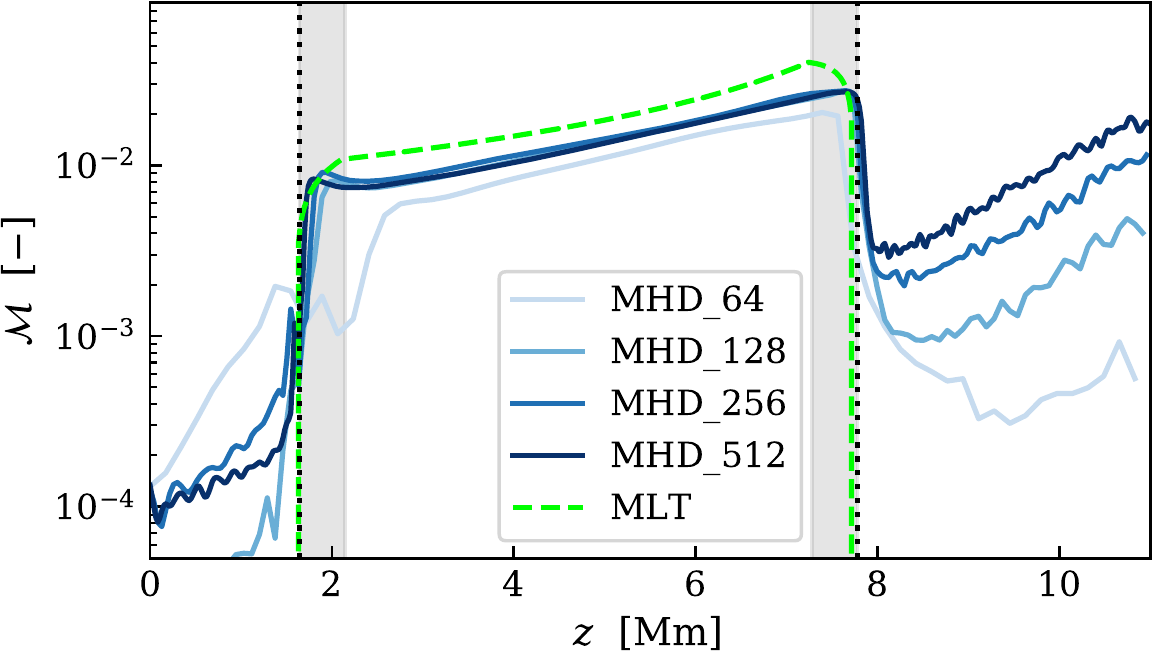}}
	\caption{Comparison between Mach number profiles derived from the MHD simulations and from mixing-length theory (MLT) with $\alpha_{\rm MLT} = 1.0$. Black dotted lines denote the boundaries of the convection zone, while the gray areas represent the artificial heating (\textit{left}) and cooling (\textit{right}) regions.}
	\label{fig:properties_mlt}
\end{figure}
\begin{figure}
	\centering
	\resizebox{\hsize}{!}{\includegraphics{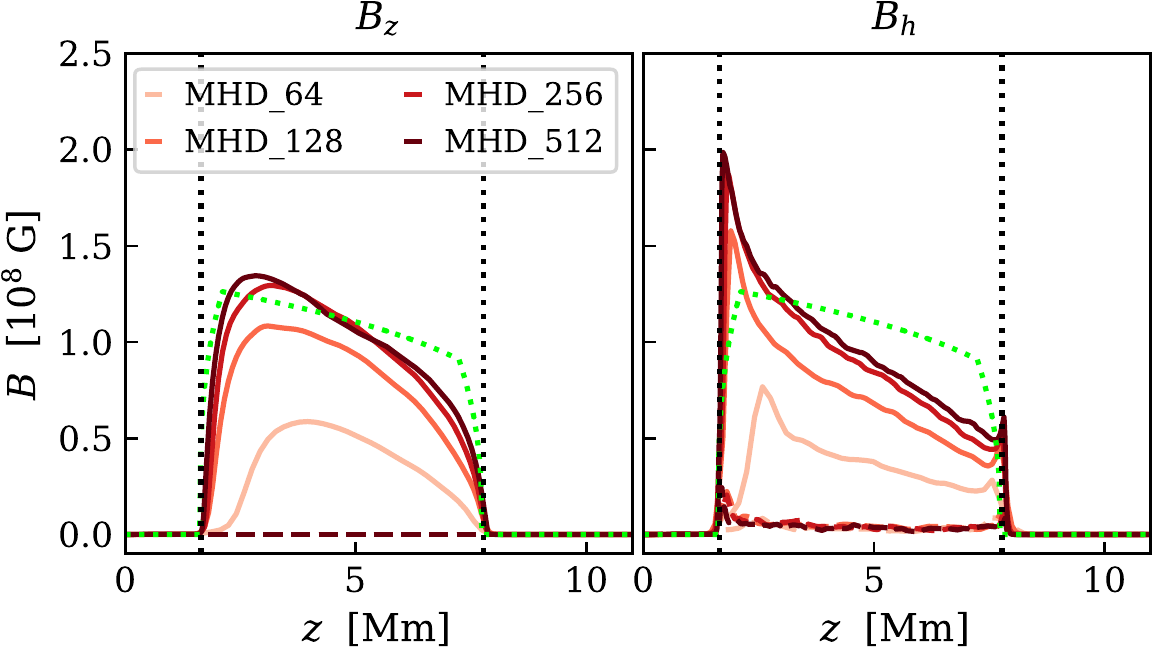}}
	\caption{Vertical profiles of the vertical (\textit{left}) and horizontal (\textit{right}) components of the magnetic field for the four MHD simulations. The rms profiles are shown in continuous lines, while dashed lines represent the mean profiles. Dotted vertical lines denote the boundaries of the convection zone. The MLT prediction for the magnetic field strength according to Eq.\,(\ref{eq:MLT_magneticfield}) and with $\alpha_{\rm B} = 0.36$ is shown in green.}
	\label{fig:properties_profiles_B}
\end{figure}
\begin{figure}
	\centering
	\resizebox{\hsize}{!}{\includegraphics{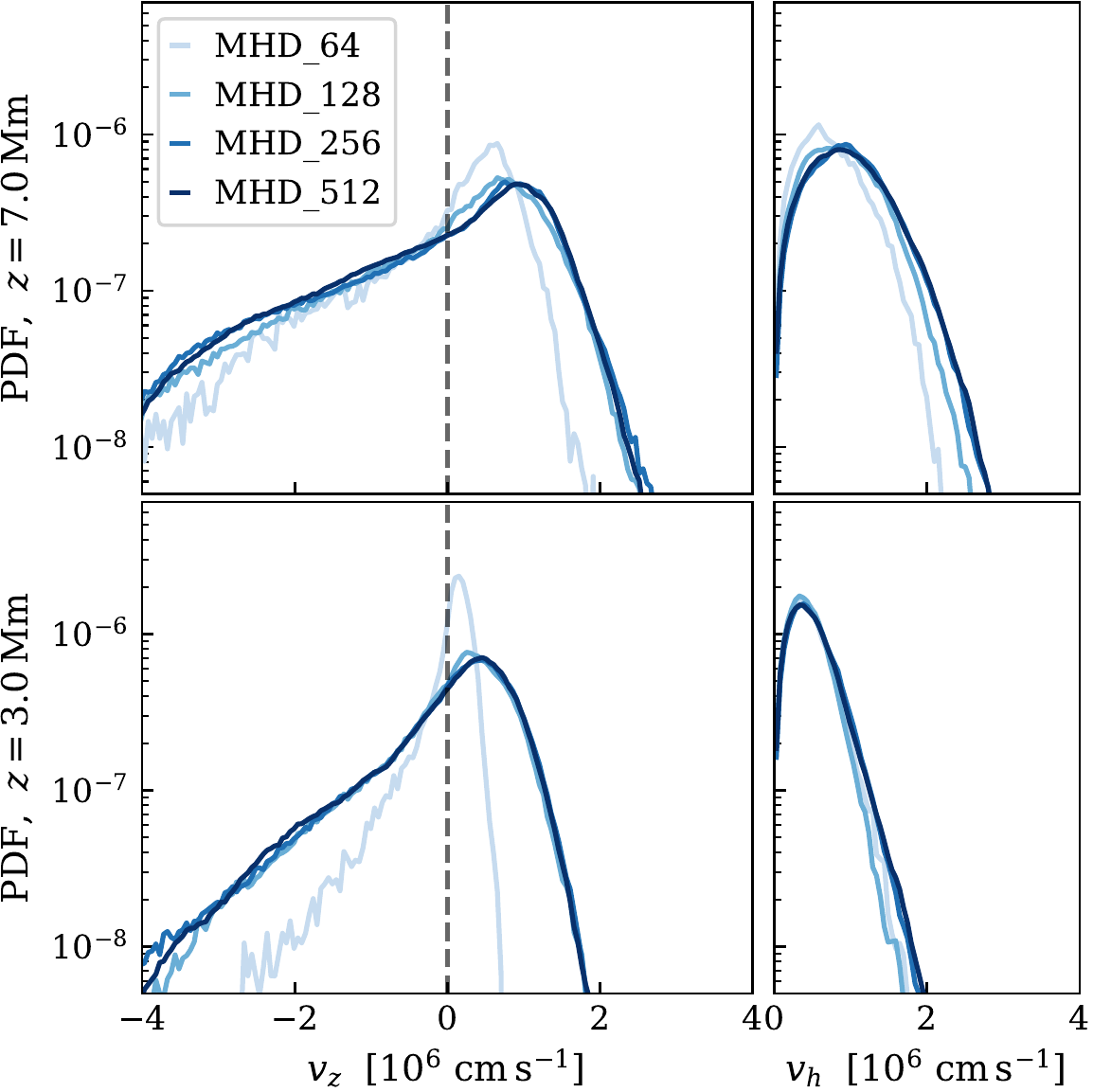}}
	\caption{Probability density functions (PDF) of the vertical (\textit{left column}) and horizontal (\textit{right column}) components of the velocity field in the four MHD simulations after reaching the quasi-steady state. The PDFs are computed over a horizontal section at $z=7.0\,{\rm Mm}$ (\textit{top row}) and at $z=3.0\,{\rm Mm}$ (\textit{bottom row}), corresponding to top and bottom convection zone, respectively. }
	\label{fig:properties_pdf_velocity}
\end{figure}
\begin{figure}
	\centering
	\resizebox{\hsize}{!}{\includegraphics{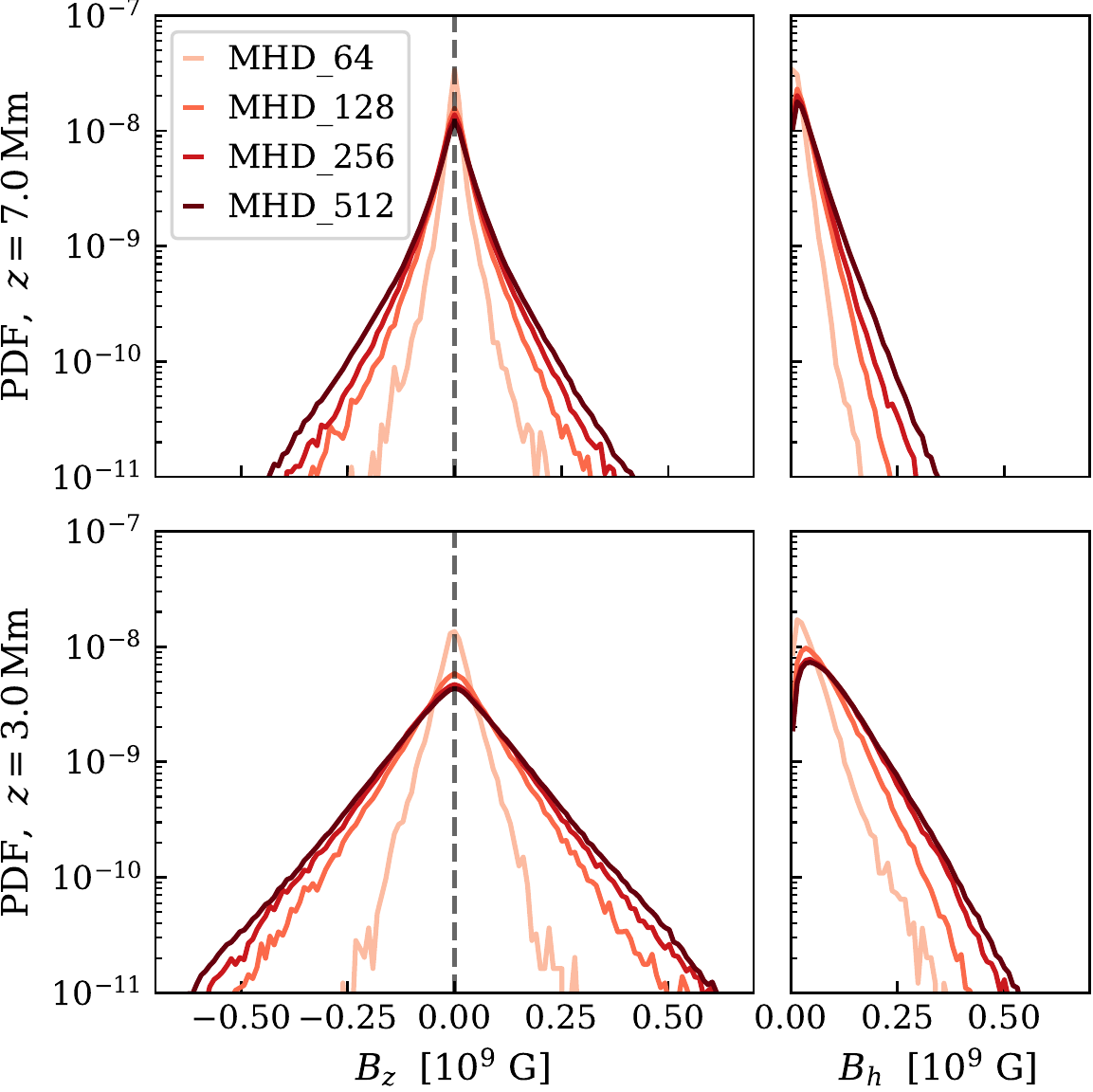}}
	\caption{Probability density functions (PDF) of the vertical (\textit{left column}) and horizontal (\textit{right column}) components of the magnetic field in the four MHD simulations after reaching the quasi-steady state. The PDFs are computed over a section at $z=7.0\,{\rm Mm}$ (\textit{top row}) and at $z=3.0\,{\rm Mm}$ (\textit{bottom row}), corresponding to top and bottom convection zone, respectively. }
	\label{fig:properties_pdf_magnetic}
\end{figure}

Once the simulations attain a quasi-steady magneto-convectional state, we can infer statistical properties of the convective flows and magnetic fields. In this section, all the results for each different simulation represent averages over all snapshots where the saturation phase has been reached. In particular, we consider the last $7$ snapshots for the $\texttt{MHD\_64}$ simulation, the last $10$ for $\texttt{MHD\_128}$, $7$ for $\texttt{MHD\_256}$, and $6$ for $\texttt{MHD\_512}$.

From a MLT perspective and assuming isotropical turbulence, we can predict the vertical profile of the one-dimensional velocity dispersion $\sigma^{\rm 1D}_{\rm MLT}$ \citep[see, e.g.,][]{1992..USB..S} as, 
\begin{equation}
    \sigma^{\rm 1D}_{\rm MLT} = \left(\frac{2 ( \gamma-1 ) \alpha_{\rm MLT} Q_{\rm T}}{\gamma \bar{\rho}} \right)^{1/3} \, , \label{eq:MLT_sigma} 
\end{equation}
where $\alpha_{\rm MLT}$ is the mixing-length parameter\footnote{The mixing-length parameter $\alpha$ is defined as the ratio between the turbulent eddies mixing-length and the local pressure scale height.} which we fix to $\alpha_{\rm MLT}=1.0$, $Q_{\rm T}$ is the convective flux, which we assume to be equal to the stellar energy flux (see Sect.\,\ref{subsec:setup}), and $\bar{\rho}$ is the mean density profile. Assuming quasi-equipartition, we can also predict the vertical profile of the one-dimensional magnetic field strength as,
\begin{equation}
    B^{\rm 1D}_{\rm MLT} = \alpha_{\rm B}\sqrt{4\pi\bar{\rho}}\, \sigma^{\rm 1D}_{\rm MLT}\,, \label{eq:MLT_magneticfield}
\end{equation}
where $\alpha_{\rm B} = B_{\rm rms}/B_{\rm eq}$ is the ratio between the rms magnetic field strength and the equipartition value found in the MHD simulations and shown in Tab.\,\ref{tab:equipartition_B}.

Figure \ref{fig:properties_profiles_v} shows the vertical profiles of the rms and mean components of the vertical and horizontal velocities, defined as $v_z$ and $v_h = \frac{1}{\sqrt{2}}\sqrt{v_x^2 + v_y^2}$, respectively. 
The rms vertical velocity grows as we approach the top of the convection zone, as predicted by MLT since the density profile decreases with $z$. The maximal amplitude of the rms profile is reached just before the artificial cooling layer. Near the two boundaries and in the bottom stable region, the rms vertical velocity amplitude is very weak, while in the top stable layer we see the imprints of vertical oscillations. On the other hand, the rms horizontal velocity profile peaks near the boundaries of the convection zone. At the top, the uprising convective plumes hit the top stable layer and are cooled down by the artificial cooling. Therefore, the plasma is pushed horizontally towards the intergranular downflows, where it begins its descent. In the deep convection zone, the descending cool plasma encounters the bottom stable stratification and the artificial heating region, thus it gets heated up and it is channeled horizontally into the convective cells. Moreover, we observe large amplitudes of horizontal velocities (in both rms and mean components) in the top stable layer. These are also imprints of waves propagating horizontally in the stable stratification. 

Qualitatively, the velocity profiles are comparable to what \citet{2013ApJ...769....1V} 
obtained for the convective envelope of a $5 M_{\odot}$ red giant and  \citet{2008ApJ...673..557M}, \citet{2014ApJ...786...24H}, and \citet{2015ApJ...803...42H} 
for the solar convection zone. 
Moreover, in Fig.\,\ref{fig:properties_mlt} we show the vertical profiles of the average three-dimensional Mach number, $\mathcal{M} = v_{\rm rms}/c_{\rm s}$, for each one of the MHD simulations. We compare the simulated profiles to a MLT estimate given by $\mathcal{M}_{\rm MLT} = \sigma^{\rm 3D}_{\rm MLT}/c_{\rm s} = \sqrt{3}\sigma^{\rm 1D}_{\rm MLT}/c_{\rm s}$, where $\sigma^{\rm 1D}_{\rm MLT}$ is given in Eq.\,(\ref{eq:MLT_sigma}). We can once more appreciate the extreme subsonic nature of the convective flows, with the maximum mean Mach number being in the order of $\mathcal{M} \sim 10^{-2}$ near the top of the convection zone. 
In addition, we show that our code yields satisfying results also with lower Mach numbers in Appendix \ref{app:low_mach_number_sims}. 

We find that the velocity profiles are in good agreement with MLT theory and that they converge with resolution in the convection zone. In the upper stable layer however, the Mach number profiles from the different simulations do not overlap. The growth of kinetic energy in this layer, seen also in Fig.\,\ref{fig:properties_profiles_v}, is due to trapped internal gravity waves. Ideally, these waves would be free to exit the simulation box, but since the top boundary conditions are fixed, the top stable layer acts as a resonant cavity. We will explore outflow boundary conditions with proper characteristic tracing in future work. 
Internal gravity waves can transport energy at the smallest scales in the simulation (see Sect.\,\ref{subsec:waves}), therefore the better the spatial resolution, the higher the kinetic energy that can be provided to such modes. 

Figure \ref{fig:properties_profiles_B} is the magnetic analog of Fig.\,\ref{fig:properties_profiles_v}. Just as for the velocity, we define the vertical and horizontal magnetic field components as $B_z$ and $B_h = \frac{1}{\sqrt{2}}\sqrt{B_x^2 + B_y^2}$. The rms magnetic magnetic vertical profiles are reversed with respect to the velocity ones, as the peaks are near the bottom of the convection zone. Similar profiles are found by \citet{2014ApJ...789..132R}, \citet{2014ApJ...786...24H}, and \citet{2015ApJ...803...42H} 
for magnetic fields generated by a small-scale dynamo in the solar convection zone. 
The magnetic field profiles qualitatively follow the MLT prediction and also converge with resolution, but the difference between the different simulations is more noticeable. In the top and bottom stable regions the magnetic field is essentially zero apart from two thin layers above and beneath the convection zone where, as we have seen in Fig.\,\ref{fig:dynamo_visualization}, magnetic field is transported by overshooting convective plasma. 

Figure \ref{fig:properties_pdf_velocity} shows the probability density functions (PDFs) of the vertical and horizontal components of the velocity field at two different heights. Near the top of the convection zone ($z=7.0\,{\rm Mm}$), the vertical velocity is characterized by an asymmetric PDF, with the probability peak being at low positive values and a long tail towards strong downflows. Such a PDF stems from the granular structure of the convective flow, where the large and slowly uprising plumes dominate the volume but the strongest velocities are found in the intergranular downflows.
Closer to the bottom ($z=3.0\,{\rm Mm}$), the qualitative shape of the vertical velocity PDF is similar, but the peak is closer to zero and the probability of having large negative velocities is lower. This follows from the decrease of the average rms velocities with depth due to the stellar stratification (see Fig.\,\ref{fig:properties_profiles_v} and Eq.\,\ref{eq:MLT_sigma}). The horizontal velocity PDFs appear to follow a chi distribution with two degrees of freedom at both heights, because $v_x$ and $v_y$ are Gaussian distributed. Moreover, their peaks are found at the same values of the vertical ones. Once more, the distribution in the deep convection zone is narrower because of stellar stratification. 

The PDFs of the vertical magnetic field are shown in the left panels of Fig.\,\ref{fig:properties_pdf_magnetic}. The profiles are symmetric at both heights in the convection zone, which is compatible with the origin of these magnetic fields being due to a turbulent small-scale dynamo with no preferred direction. The profiles are narrower with respect to the velocity ones, showing a higher intermittency of the magnetic fields \citep[][]{1996JFM...306..325B}. 
The tails are more extended at $z=3.0\,{\rm Mm}$ since in the deeper layers of the convection zone the mean magnetic field amplitude is larger (see Fig.\,\ref{fig:properties_profiles_B}). Similar conclusions can be inferred from the right panels of Fig.\,\ref{fig:properties_pdf_magnetic}, where we show the PDFs of the horizontal magnetic field, $B_h$. 

Both in Fig.\,\ref{fig:properties_pdf_velocity} and Fig.\,\ref{fig:properties_pdf_magnetic} the PDFs appear to converge with resolution. Moreover, qualitatively similar results have been obtained by \citet{2004ApJ...601..512B}, \citet{2008ApJ...673..557M}, \citet{2014ApJ...789..132R}, \citet{2014ApJ...786...24H}, and \citet{2015ApJ...803...42H} 
in the context of solar convection. This validates our numerical treatment of magneto-convection flows with a very different code.
%
%
\subsection{Waves} 
\label{subsec:waves}

\begin{figure*}
	\centering
	\resizebox{\hsize}{!}{\includegraphics{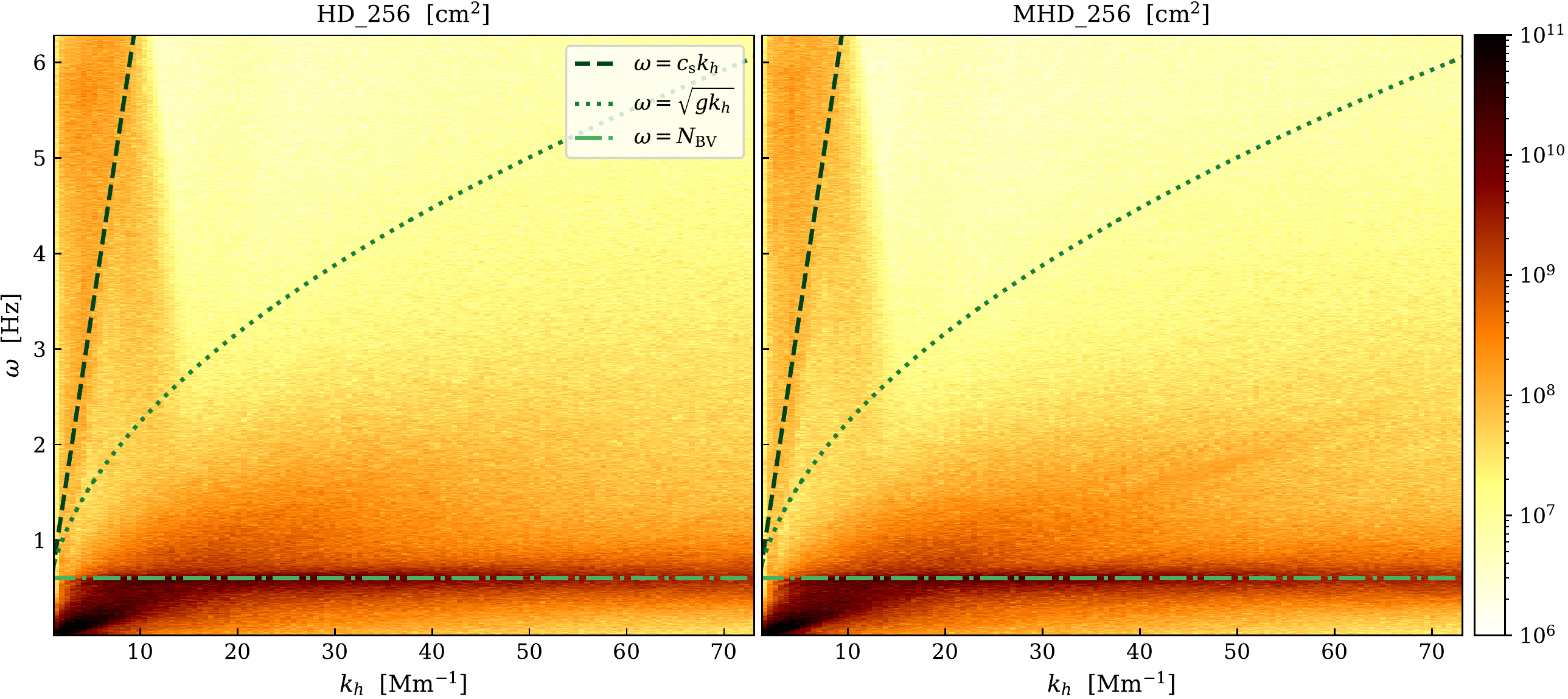}}
	\caption{Wave diagnostic diagrams, also knows as $k_h$-$\omega$ diagrams, for the $\texttt{HD\_256}$ (\textit{left}) and $\texttt{MHD\_256}$ (\textit{right}) simulations at $z=8.5\,{\rm Mm}$. The sampling rate and grid cell size are $\Delta t_{\rm s} = 0.5\,{\rm s}$ and $\Delta x = 42.0\,{\rm km}$, respectively. The strength of the signal is 
	color-coded. The dispersion relations for sound waves and surface modes are shown in dashed and dotted lines, respectively. The Brunt-Väisälä frequency is depicted by a dash-dotted line.}
	\label{fig:waves_diagnostics}
\end{figure*}
\begin{figure}
	\centering
	\resizebox{\hsize}{!}{\includegraphics{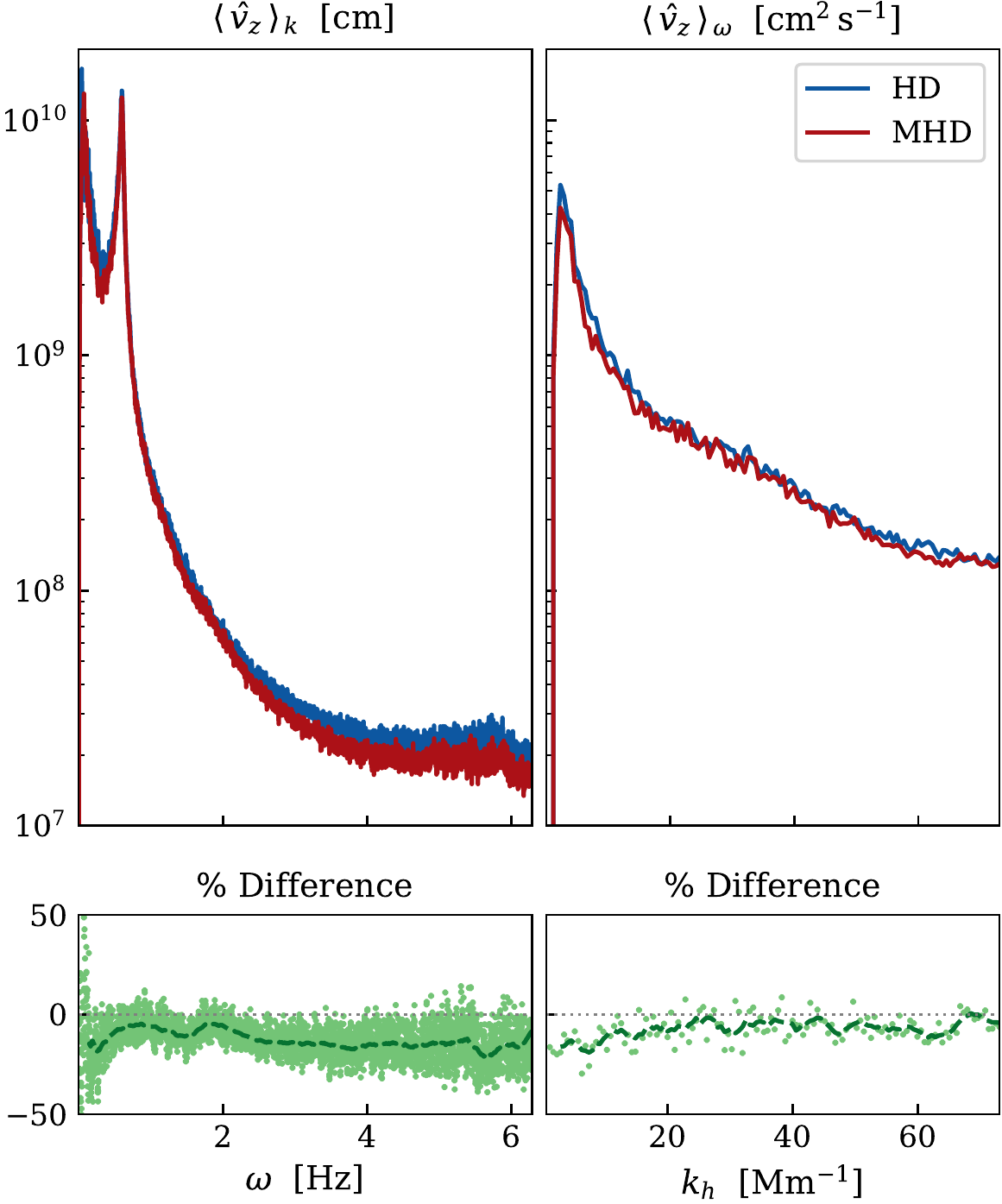}}
	\caption{Comparison between the mean power as a function of frequency $\omega$ (\textit{left}) and of horizontal wavenumber $k_h$ (\textit{right}) for the $\texttt{HD\_256}$ and $\texttt{MHD\_256}$ simulations at $z=8.5\,{\rm Mm}$. The bottom panels show the relative difference between the two signals in percentage. The green dashed line shows a smoothed interpolation of the data points.}
	\label{fig:waves_comparison}
\end{figure}

Our fully compressible approach allows us to study the excitation and propagation of acoustic (or pressure) waves, in addition to internal gravity waves.
Therefore, in order to characterize the oscillatory phenomena visible in the stable regions above and below the convection zone in both hydrodynamical and MHD simulations (see Fig.\,\ref{fig:convection_visualization} and Fig.\,\ref{fig:dynamo_visualization}), we performed a Fourier analysis on a time series of $N_{\rm s} = 9\,000$ snapshots of a horizontal section of the vertical velocity $v_z$ taken at $z=8.5\,{\rm Mm}$, that is in the top stable layer and $\sim 0.7\,{\rm Mm}$ above the convection zone. We used the simulations $\texttt{HD\_256}$ and $\texttt{MHD\_256}$ and we started the analysis at $t=5\,000\,{\rm s}$ in both cases. The sampling rate is $\Delta t_{\rm s} = 0.5\,{\rm s}$, hence we cover a frequency domain ranging from $\omega_{\rm min} = 2\pi/T = 1.40 \times 10^{-3}\,{\rm Hz}$, where $T = N_{\rm s} \Delta t_{\rm s} = 4\,500\,{\rm s}$, to the Nyquist frequency $\omega_{\rm Nyquist} = \pi/\Delta t_{\rm s} = 6.28\,{\rm Hz}$. The horizontal wavenumber instead, which we define as $k_h^2 = k_x^2 + k_y^2$, ranges from $k_{\rm h,\,min} = 5.71 \times 10^{-1}\,{\rm Mm}$ to $k_{\rm h,\,Nyquist} = 73.1\,{\rm Mm}$ given the resolution and the size of the box. Before performing a Fourier transform over the time sequence, we apodized the data by a cosine-bell function to avoid artifacts coming from the nonperiodicity of the signal. We did not need to apply the same procedure for the spatial Fourier transform since the lateral boundaries of the box are periodic. 

The results of the Fourier analysis performed over the hydrodynamical simulation \texttt{HD\_256} and MHD simulation \texttt{MHD\_256} are presented as diagnostic diagrams, or $k_h$-$\omega$ diagrams, in Fig.\,\ref{fig:waves_diagnostics}. In such diagrams, a single wave form would appear as a dark spot at the relative frequency $\omega$ and horizontal wavenumber $k_h$, while more complex oscillations and mode families can give rise to rays or ridges. In a stratified, compressible medium, the behavior of a generic wave is characterized by the local dispersion relation \citep[see, e.g.,][]{2014masu.book.....P}, 
\begin{equation}
    k_z^2 = \frac{(\omega^2  - \omega_{\rm ac}^2)}{c_{\rm s}^2} - \frac{(\omega^2 - N_{\rm BV}^2) k_h^2}{\omega^2}\,,\label{eq:local_dispersion_relation}
\end{equation}
where $k_z$ is the vertical wavenumber and $c_s$ is the sound speed. The acoustic cutoff frequency $\omega_{\rm ac}$ \citep[see, e.g.,][]{1984ARA&A..22..593D} 
and the Brunt-Väisälä frequency $N_{\rm BV}$ are defined as,
\begin{equation}
    \omega_{\rm ac}^2 = \frac{c_{\rm s}^2}{4H_{\rho}^2}\left(1 - 2\frac{d H_{\rho}}{dz}\right)\,,
\end{equation}
\begin{equation}
    N_{\rm BV}^2 = g\left( \frac{1}{H_{\rho}} - \frac{1}{\gamma H_p}\right)\,,
\end{equation}
where $H_{\rho}$ and $H_p$ are the density and pressure scale heights, respectively.
The solutions of setting $k_z^2=0$ in Eq.\,(\ref{eq:local_dispersion_relation}) separate two domains where waves can propagate ($k_z^2 > 0$) and one region where waves are evanescent ($k_z^2 < 0$). For $k_h > 1/(2 H_{\rho})$, the boundaries in the frequency domain between propagating and evanescent regions can be approximated by the Brunt-Väisälä frequency at the lower boundary and by the dispersion relation for sound waves in a homogeneous, compressible medium $\omega = c_{\rm s} k_h$ for the upper boundary. In our case, $1/(2 H_{\rho}) \sim 0.5\,{\rm Mm}^{-1}$, so this approximation is valid for most of the parameter space displayed in Fig.\,\ref{fig:waves_diagnostics}. In the same figure, we show these two relations as dash-dotted and dashed lines, respectively, where $c_{\rm s}$ and $N_{\rm BV}$ are computed using our equilibrium profiles. Moreover, we also show the dispersion relation for surface gravity waves ($f$-modes), $\omega = \sqrt{g k_h}$, with a dotted line. For more details on the local dispersion relation of waves in a gravitationally stratified background, the reader can refer to \citet[][]{2017ApJ...835..148V}. 

The two panels of Fig.\,\ref{fig:waves_diagnostics} appear very similar qualitatively. In both cases, we observe several different modal components. Gravity-modified pressure waves ($p$-modes) are responsible for the signal in the top left of both panels, that is above the $\omega = c_{\rm s} k_h$ relation. At the top of the diagram we observe a reflection of the signal due to an aliasing phenomenon, also reported by \citet[][]{2006ApJ...642.1057H}. 
Individual ridges due to internal gravity waves ($g$-modes) can be observed at small horizontal wavenumbers ($k_h \lesssim 3\,{\rm Mm}^{-1}$) and just below the Brunt-Väisälä frequency, while at smaller scales they cannot be resolved individually and give rise to a horizontal band along $\omega = N_{\rm BV}$. The $g$-modes are excited by the convective plumes perturbing the stratified plasma in the stable region, while $p$-modes are mainly produced and propagate within the convective region.
The result is a horizontal flow with an oscillatory component ($g$-modes), that resemble waves at the surface of the see, and the propagation of an over-pressure in the stable layers ($p$-modes). The turbulent convective noise is responsible for the strong smeared signal in the bottom left of both panels which extends through the evanescent zone. This shows once more that the convective flow is able to overshoot into the stable layers and there perturb vertically the plasma. Similar results have been obtained by \citet[][]{2006ApJ...642.1057H}  
for the same hydrodynamical setup as used here and by \citet[][]{2007ApJ...667..448M} 
for an oxygen shell burning $23\,M_{\odot}$ stellar model. 

To better expose the difference between the two plots of Fig.\,\ref{fig:waves_diagnostics}, we show in Fig.\,\ref{fig:waves_comparison} the power averaged over the horizontal wavenumber $k_h$, $\langle \hat{v}_z \rangle_k$ (\textit{left}), and over the frequency $\omega$, $\langle \hat{v}_z \rangle_{\omega}$ (\textit{right}), for both simulations. The curves are in both panels qualitatively similar, but the hydrodynamical simulation display slightly higher values. The two peaks at low frequency ($\omega \sim 0.1 \,{\rm Hz}$ and $\omega \sim 1 \,{\rm Hz}$) correspond to the convective noise signal and to the $g$-mode ridge visible in Fig.\,\ref{fig:waves_diagnostics}. The local crest at high frequency ($\omega \sim 5.5 \,{\rm Hz}$) corresponds to the $p$-modes signal, while the single peak at low horizontal wavenumber ($k_h \sim 2 {\rm Mm}^{-1}$) is mainly due to the convective noise signal.

The bottom panels of Fig.\,\ref{fig:waves_comparison} show the relative difference between the two curves computed as,
\begin{equation}
\text{Difference} = \frac{\langle \hat{v}_z \rangle^{\rm MHD}_i - \langle \hat{v}_z \rangle^{\rm HD}_i}{\langle \hat{v}_z \rangle^{\rm HD}_i}\,, \label{eq:relative_diff_power} 
\end{equation}
where $i = k_h,\,\omega$, respectively. We see that waves in the \texttt{MHD\_256} simulation have, in average, less power than in \texttt{HD\_256} for all frequencies and horizontal wavenumbers. 
In total, the relative difference between the two is $-11.5\,\%$. By using the Brunt-Väisäla frequency and the sound waves dispersion relation to separate the different contributions, we identify a deficit of $13.1\%$ in power related to internal gravity waves and $16.3\%$ to acoustic waves. Moreover, the relative difference varies with $\omega$ and $k_h$. We can identify two dips of $\sim-20\,\%$ in the left plot: one at very low and another one at large frequency. The first one ($\omega \sim 0.1\,{\rm Hz}$) is related to the convective noise signal, which appear to be stronger in the hydrodynamical simulation, while the second one ($\omega \sim 5.5\,{\rm Hz}$) refers to the $p$-modes signal. If the first dip can be easily explained with the difference of convective kinetic energy between the two simulations due to the feedback of magnetic fields (see Fig.\,\ref{fig:dynamo_PS_evolution} and Fig.\,\ref{fig:dynamo_PS_comparison}), the deficit associated with p-modes is not necessarily expected. Although the power associated with pressure waves scales with the eight power of the characteristic turbulent velocity \citep{1952RSPSA.211..564L}, 
magneto-acoustic waves can be excited by means of Lorentz forces. Therefore, the loss in power due to a lower turbulent velocity could be compensated by the action of the magnetic fields. 
The two smallest differences are found in correspondence with the peak of $g$-modes ($\omega \sim 1\,{\rm Hz}$) and around $\omega \sim 2\,{\rm Hz}$, where the $\texttt{MHD}$ run display a small bump associated with a slightly stronger signal in the evanescent region. This signal is mysterious in origin and we speculate could be caused by the magnetic fields in the top convection zone. 

A detailed analysis of the causes behind these differences are beyond the scope of this paper. However, there are clear hints indicating the impact of magnetic fields on the generation and propagation of wave modes in the upper stable layer.     
%
%
\section{Summary and conclusions}
\label{sec:conclusion}
In this paper we presented a proof of concept for three-dimensional, fully compressible, numerical simulations of magneto-convection in a Cartesian grid with the RAMSES code. In order to deal with low-Mach number and low-amplitude fluctuations typical of convective motions, we adapted the numerical solvers of RAMSES by implementing a well-balanced scheme. 

We ran hydrodynamical and MHD simulations of a He shell flash convective envelope model with different resolutions, ranging from $N=64^3$ to $N = 512^3$. A turbulent convection zone naturally develops from random perturbations in the initial density profile, and the emerging rms velocity field is compatible with MLT predictions.
In the Appendix \ref{app:low_mach_number_sims} we show that our code is able to deal with flows characterized by Mach numbers as low as $\mathcal{M} \sim 10^{-3}$. We note that, in principle, our well-balanced scheme allows to model even lower Mach number flows. The only limitation regards the computational cost. Indeed, as we lower the characteristic convective velocity of the simulation, we lower the time-step as well because of the Courant condition.
Regarding the MHD simulations, a turbulent small-scale dynamo effectively amplifies a weak seed magnetic field up to $\sim 18-37\,\%$ of the equipartition value depending on the resolution. Once saturated, the magnetic fields dominate the dynamics of the convection zone at small scales. In the two stable layers, we saw indications of convective overshoot and we identified propagating pressure and internal gravity waves. 

The resulting properties of the convection region and of the magnetic fields converge by increasing the simulation resolution, as we can see from the power spectra analysis (Figs.\,\ref{fig:convection_PS_comparison} and \ref{fig:dynamo_PS_comparison}), from the vertical profiles (Figs.\,\ref{fig:properties_profiles_v} and \ref{fig:properties_profiles_B}), and from the PDF distributions (Figs.\,\ref{fig:properties_pdf_velocity} and \ref{fig:properties_pdf_magnetic}) of velocity and magnetic fields. In particular, we observe an almost statistically equivalent result for the velocity field for the resolutions $N=128^3,\,256^3,$ and $512^3$, while the difference is more pronounced regarding the magnetic fields. In both cases, the low-resolution run ($N=64^3$) presents important departures from the expected results and it is therefore insufficient to capture the magneto-convective dynamics of this setup. Indeed, the low-resolution run resolves the minimum pressure scale heights ( $H_{\rm p}^{\rm min}\sim0.7\,{\rm Mm}$ ) with only $\sim 4$-$5$ grid-cells, which is not sufficient for an (magneto-)hydrodynamical code. We recall that RAMSES employs a second-order numerical scheme in both hydrodynamical and MHD solvers. Using a higher-order scheme could improve the results even at low-resolutions. Only for the velocity field in the top stable layer we do not observe a convergence with resolution (see Figs.\,\ref{fig:properties_profiles_v} and \ref{fig:properties_mlt}). There, the fixed top boundary condition creates a resonant cavity which traps the waves generated at the interface with the convection zone. A more realistic treatment of the boundary conditions will address this problems in future works.

Qualitatively, we find similar results to several other studies on convection and magneto-convection. Our results differ from the two-dimensional simulations obtained by \citet[][]{2006ApJ...642.1057H} 
with the same initial model. The reason for this discrepancy is the artificial cooling layer that we introduced at the top of the convection zone to mimic radiative losses. Indeed, the cooling layer causes the intergranular downflows which drive the convective motions in our simulations (see Fig.\,\ref{fig:convection_visualization}), whereas in \citet[][]{2006ApJ...642.1057H} 
convection is induced by hot upflow fingers generated in the heated bottom layers. This explains why our results are more similar to surface solar convection simulations \citep[see, e.g.,][]{2014ApJ...789..132R, 2014ApJ...786...24H, 2015ApJ...803...42H} 
than to core convection simulations \citep[see, e.g.,][]{2004ApJ...601..512B, 2005ApJ...629..461B}. 
In any case, our goal was not to reproduce realistic results regarding a particular stellar convection scenario, but to demonstrate the feasibility of simulating three-dimensional compressible magneto-convection with the RAMSES code, and in that respect we consider the proof of concept successful.

The analysis carried out with respect to the propagation of oscillations in the stable region above the convection zone revealed a deficit of power in the magneto-hydrodynamical simulation. In particular, because of our fully compressible approach, we could measure a deficit of $16.3\%$ in power associated with pressure waves. This difference with respect to the purely hydrodynamical case cannot be readily explained by the reduction of convective kinetic energy due to a small-scale dynamo action in the MHD simulation. Indeed, magnetic fields amplified by the same small-scale dynamo  can also be sources of magneto-acoustic waves. Therefore, we provide clear indications that magnetic fields play a non trivial and important role in the excitation and propagation of acoustic waves in stellar interiors.

In the future, we aim to expanding our methodology to global models of stellar magneto-convection. In particular, we are interested in the propagation of waves within stars. Our compressible approach, similarly to \citet[][]{2020A&A...641A..18H}, 
allows for the correct treatment of both pressure and internal gravity modes excited by the convective motions. 
However, the coupling between magnetic fields and these oscillations is of particular interest for asteroseismological observations. For example, the magnetic greenhouse effect could trap wave-modes in the core and therefore suppress dipole oscillations \citep[][]{2015Sci...350..423F, 2016Natur.529..364S}. 
Magneto-hydrodynamical numerical simulations of fully compressible convection, such as the ones presented in this paper, represent a viable solution to study these phenomena. 

The price to pay though is the high computational cost required. Indeed, the $\texttt{MHD\_512}$ simulation required $2.5 \times 10^6\,{\rm core}\,{\rm h}$ with $4\,608$ cores to run for $6\,300\,{\rm s}$ physical time, while for the $\texttt{MHD\_256}$ simulation we used $1\,152$ cores and a total of $1.7 \times 10^5\,{\rm core}\,{\rm h}$ to simulate $9\,000\,{\rm s}$ physical time. That corresponds approximately to $6$ and $9$ convective turnovers times, respectively. \citet[][]{2020A&A...641A..18H} 
estimate a need of $44 \times 10^6\,{\rm core}\,{\rm h}$ to simulate a full $3\,M_{\odot}$ stellar model for $700\,{\rm h}$ physical time ($\sim 17$ convective turnover times) on a spherical grid of size $960\,(r) \times 360\,(\vartheta) \times 720\,(\varphi)$ with the compressible Seven-League Hydro (SLH) code \citep[][]{2013PhDThesis..M}. 
Our MHD simulations on a Cartesian grid would increase the computational time cost required, because of the higher number of degrees of freedom and because of the higher resolution needed given the geometry of the grid. However, even taking into account the extra costs, we estimate our numerical simulations to be rather competitive. 
Moreover, the AMR capabilities of the RAMSES code could help to reduce this gap. Another possibility is to alleviate the stringent time-step constraints imposed by the highly subsonic convective flow using the reduced speed of sound technique \citep[][]{2005ApJ...622.1320R, 2015ApJ...803...42H}. 

An area of improvement concerns energy conservation. As stated in Sect.\,(\ref{subsec:well-balances_scheme}), we solve for the conservation of specific entropy instead of total energy, since entropy is the fundamental quantity governing the dynamics of convectively unstable systems. However, this choice implies that energy conservation is not ensured by the numerical scheme. A direct consequence is that the numerical energy flux in the convection zone is larger than the theoretical one by tens of percent. We believe that, in the future, this problem can be addressed by employing higher-order schemes to solve for the total energy conservation and at the same time accurately capturing the dynamics of low amplitude entropy perturbations. Moreover, we plan to improve our simulation setup by avoiding discontinuities in the heating and cooling functions as well as in the equilibrium profiles. These discontinuities are sources of truncation errors that can lead to spurious energy fluxes.

In conclusion, we proved the feasibility of three-dimensional, fully compressible numerical simulations of magneto-convection in the low-Mach number regime with the RAMSES code. This allowed us to study the self-consistent amplification of magnetic fields in stellar convection zones together with their interaction with the generation and propagation of pressure and internal gravity waves in the adjacent stellar cores, radiative zones, or stellar atmospheres.   
%
%

\begin{acknowledgements}
JRRC acknowledges support by SNF under grant ID 200020\_182094. The simulations have been run on the Piz Daint and Eiger machines at the Swiss National Supercomputer Centre (CSCS) under projects uzh5, uzh3, and s1006. We would like to sincerely thank the anonymous referee, O. Steiner, G. Vigeesh, and E. Quataert for very enlightening discussions and comments.
\end{acknowledgements}


%
%

\bibliographystyle{aa} 
\bibliography{biblio.bib} 

%
%

\begin{appendix}

%
%
\section{Low-Mach number simulations}
\label{app:low_mach_number_sims}
\begin{figure}[h]
	\centering
	\resizebox{\hsize}{!}{\includegraphics{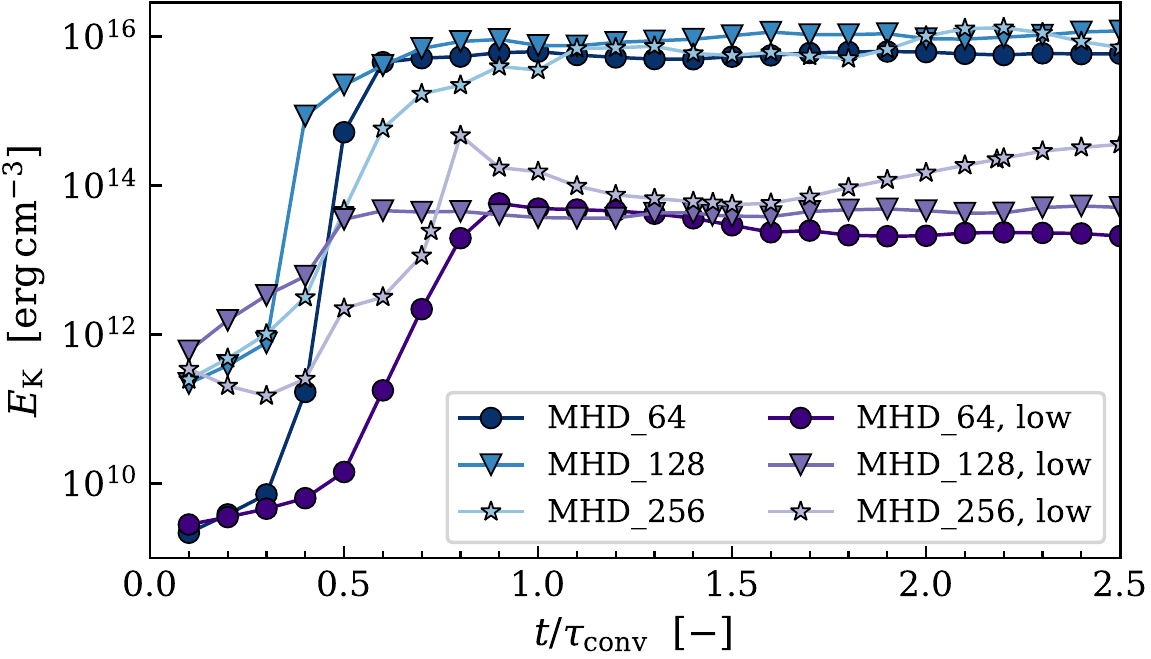}}
	\caption{
	Time evolution of the mean turbulent kinetic energy density $E_{\rm K}$ in the convection zone for simulations with nominal ($L = 3.21 \times 10^7\,L_{\odot}$, blue) and low ($L = 3.21 \times 10^4\,L_{\odot}$, purple) stellar luminosities. The convective time scales are assumed to be $\tau_{\rm conv} = 1\,000\,{\rm s}$ for the nominal simulations (see Eq.\,\ref{eq:convective_time}) and $\tau_{\rm conv} = 10\,000\,{\rm s}$ for the low luminosity ones. 
	}
	\label{fig:app_energy}
\end{figure}
\begin{figure}[h]
	\centering
	\resizebox{\hsize}{!}{\includegraphics{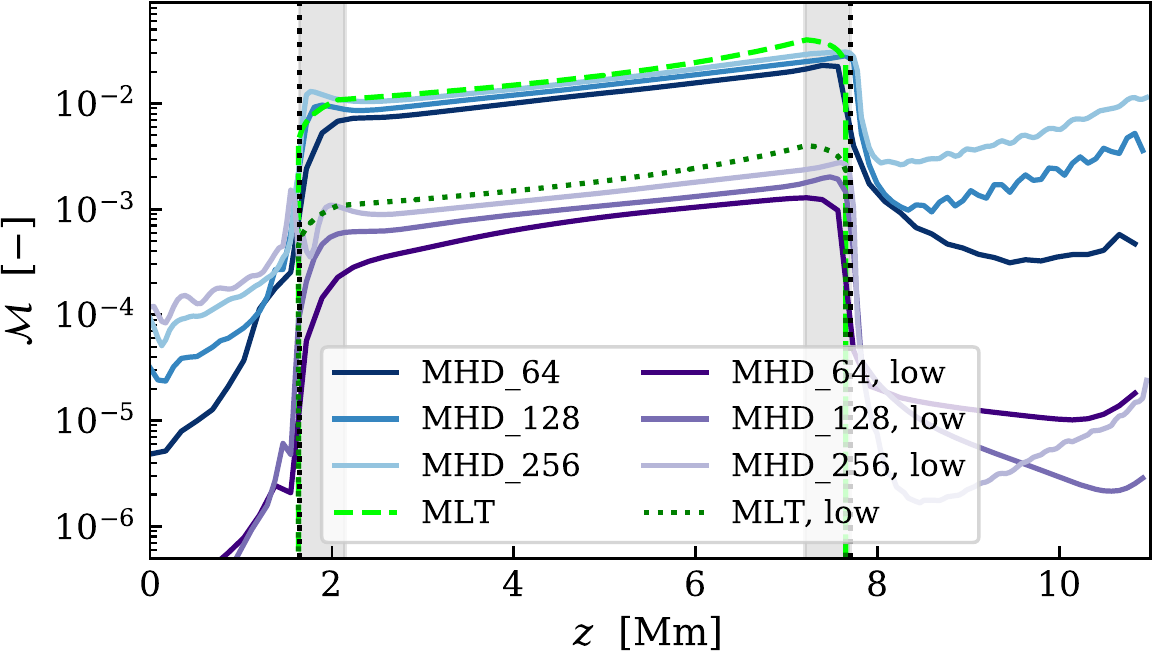}}
	\caption{
	Mach number profiles derived from the nominal and low luminosity simulations. MLT predictions with $\alpha_{\rm MLT} = 1.0$ are shown for both scenarios. Black dotted lines denote the boundaries of the convection zone, while the gray areas represent the artificial heating (\textit{left}) and cooling (\textit{right}) regions.
	}
	\label{fig:app_mlt}
\end{figure}

In this appendix, we demonstrate that our fully compressible approach is able to deal with low-Mach number flows. For this purpose, we lower the volumetric heating and cooling rates of the model presented in Sect.\,\ref{subsec:setup}. Doing so, we reduce the stellar energy flux and luminosity, and consequently also the characteristic convective velocity and typical Mach number. 

We adjusted the volumetric heating and cooling rates so that the stellar luminosity reaches $L = 3.21\times10^4\,L_{\rm \odot}$, which is $10^3$ times lower than the nominal value used in this paper. According to MLT theory and Eq.\,(\ref{eq:MLT_sigma}), the characteristic convective velocity is reduced by a factor $10$. We ran magneto-hydrodynamical simulations with both nominal and low luminosity setups with resolutions N = $64^3,\,128^3$, and $256^3$. We also reduced the amplitude of the initial random density perturbations $\delta \rho$ to $\delta \rho /\rho^{\rm eq} \sim 10^{-5}$. We ran the different simulations for $2.5$ convective time scales, which corresponds to $2\,500\,{\rm s}$ physical time for the nominal luminosity simulations and $25\,000\,{\rm s}$ for the low luminosity ones.  

Figure \ref{fig:app_energy} shows the time evolution of the mean turbulent kinetic energy for both nominal and low luminosity simulations. We see that the nominal luminosity runs quickly reach a quasi-steady state around $E_{\rm K} \sim 10^{16}\,{\rm erg}\,{\rm cm}^{-3}$, just as in Fig.\,\ref{fig:convection_energy}. Therefore, the amplitude of the initial perturbations does not affect the properties of convection. The low luminosity simulations instead all stabilize around $E_{\rm K} \sim 10^{14}\,{\rm erg}\,{\rm cm}^{-3}$. This value validates the MLT prediction of a reduction of convective velocities by a factor $\sim 10$. 

In Fig.\,\ref{fig:app_mlt} we plot the mean Mach number profiles for the six simulations. We used outputs from the last $1.5\,\tau_{\rm conv}$ of each simulation to compute the mean profiles, so that the quasi-steady state is already attained. The nominal luminosity ones are very similar to the ones shown in Fig.\,\ref{fig:properties_mlt}, while the low ones only reach $\mathcal{M} \sim 10^{-3}$ in the convection zone. In both cases the simulations are in good agreement with the MLT theory and the results converge with resolution in the convection zone. 

In conclusion, we proved that the well-balanced version of the RAMSES code presented in this paper is able to simulate low-Mach number (magneto-)convection, down to $\mathcal{M} \sim 10^{-3}$. Moreover, we have shown once more the good agreement between our numerical simulations and MLT predictions.
%

%
%
%

\end{appendix}

\end{document}